\documentclass{article}

\usepackage{PRIMEarxiv}

\usepackage[dvipsnames]{xcolor}
\usepackage{graphicx}
\usepackage{amsmath}
\usepackage{amssymb}
\usepackage{booktabs}
\usepackage[hyphens]{url}
\usepackage[format=plain,labelformat=simple,labelsep=period,font=small]{caption}

\usepackage[utf8]{inputenc}
\usepackage[T1]{fontenc}
\usepackage{bm}
\usepackage{microtype}
\graphicspath{{figure/}}

\definecolor{linkblue}{rgb}{0.21,0.49,0.74}
\usepackage[breaklinks,colorlinks,allcolors=linkblue]{hyperref}

\hypersetup{pdftitle={PIM: A Physical Intuition Model for Low-Latency Prediction of Object Interaction Outcomes}}

\title{PIM: A Physical Intuition Model for Low-Latency Prediction of Object Interaction Outcomes}

\author{ Tangwen Luo \\
  State Key Laboratory of CAD \& CG\\
  Zhejiang University\\
  Hangzhou, Zhejiang, China\\
   \And
  Chuankun Zheng \\
  State Key Laboratory of CAD \& CG\\
  Zhejiang University\\
  Hangzhou, Zhejiang, China\\
    \And
  Yuchi Huo \\
  State Key Laboratory of CAD \& CG\\
  Zhejiang University\\
  Hangzhou, Zhejiang, China\\}

\begin{document}
\maketitle
\begin{abstract}
Predicting object interaction outcomes from initial conditions is a central problem in data-driven physical reasoning. Existing methods often use rigid-body-level states or predict the evolution through autoregressive rollouts. Coarse representations cannot represent local deformation, while rollout errors may compound over long prediction horizons. Instead, we introduce PIM, a Physical Intuition Model that directly predicts a physical system's equilibrium state from its initial mesh geometry, kinematics, and physical attributes. PIM combines local k-nearest-neighbor (KNN) attention to encode intra-object geometry, anchor-based cross-object attention to model interactions, and a dual-branch decoder that predicts object-level translations and per-vertex residual displacements. Across MuJoCo scenarios involving both rigid and deformable interactions, PIM achieves lower vertex and centroid errors than the evaluated baselines. By predicting complete equilibrium states in a single forward pass, PIM substantially reduces latency relative to autoregressive learned models and MuJoCo rollouts under the evaluated protocols, enabling low-latency inference when only the final outcome is required.
\end{abstract}

\section{Introduction}
Simulating how objects move, collide, and deform is important in computer graphics, robotics, and embodied intelligence. Conventional physics simulators can resolve collision, contact, friction, and deformation with high fidelity, but numerical time stepping can be computationally expensive when generating large datasets or simulating complex scenes. Learning-based models offer an alternative approach to predicting object interaction outcomes from initial conditions. Adopting the terminology of prior work on predicting block tower state~\cite{lerer2016physical}, we use \emph{physical intuition} to denote the ability to infer the equilibrium state of interacting objects from their initial geometry, kinematics, and physical attributes.

Existing physical AI models use object-, particle-, graph-, or mesh-based representations primarily for stepwise dynamics prediction~\cite{battaglia2016interactionnet,li2019dpinet,sanchez2020gns,pfaff2021meshgnn,allen2023fignet}. Instead, we present PIM, a Physical Intuition Model that takes the initial scene state as input and directly predicts the equilibrium positions of all mesh vertices. This representation accommodates both rigid-body motion and non-rigid deformation, but requires the model to jointly reason about object geometry, physical attributes, and inter-object interactions across diverse contact phenomena. Throughout this paper, \emph{equilibrium state} denotes a stationary configuration in approximate static equilibrium, reached after transient motion has decayed and the net forces and torques are balanced up to numerical tolerance. States involving persistent motion, including dynamic equilibria, are outside the scope of this work.

In this work, PIM aims for object interactions. Each object is represented by vertex tokens augmented with object-level and scene-level conditions. Local KNN self-attention captures intra-object geometry, anchor-based cross-object attention exchanges information between objects, and a dual-branch decoder separates center displacement from vertex-level residual motion. PIM therefore predicts both global object movement and local geometric change within a unified mesh representation.

We evaluate PIM on four controlled MuJoCo scenarios spanning planar and irregular-surface collisions, structural collapse, and rigid--deformable interaction. Comparisons with direct equilibrium-state predictors and autoregressive rollout baselines, together with controlled ablations and runtime measurements, demonstrate lower equilibrium-state errors and low-latency inference under the evaluated protocols.

The main contributions of this work are:
\begin{itemize}
    \item We present a unified physical intuition model that directly predicts object interaction outcomes from initial geometry, kinematic conditions, and physical attributes. By representing these outcomes as equilibrium meshes, the formulation jointly accounts for rigid-body motion and localized deformation while avoiding autoregressive rollouts when only the final interaction outcome is required.

    \item We introduce PIM, which takes a vertex-based representation as input and combines local KNN self-attention to encode intra-object geometry with cross-object attention over source anchors selected via farthest point sampling (FPS). The latter enables information exchange between objects through a fixed number of source tokens. A dual-branch decoder then decomposes each object’s terminal vertex displacement into an object-level centroid translation and zero-mean per-vertex residuals, preventing redundant encoding of global motion while preserving the full expressiveness of the displacement field.
    
    \item We develop a composite loss that supervises terminal vertex coordinates, object centroids, and edge lengths. These loss terms guide vertex-wise motion, global placement, and local final geometry, respectively, allowing PIM to learn non-rigid deformation without imposing preservation of the initial shape.
    
\end{itemize}

\begin{figure*}[!htbp]
    \centering
    \includegraphics[width=\textwidth]{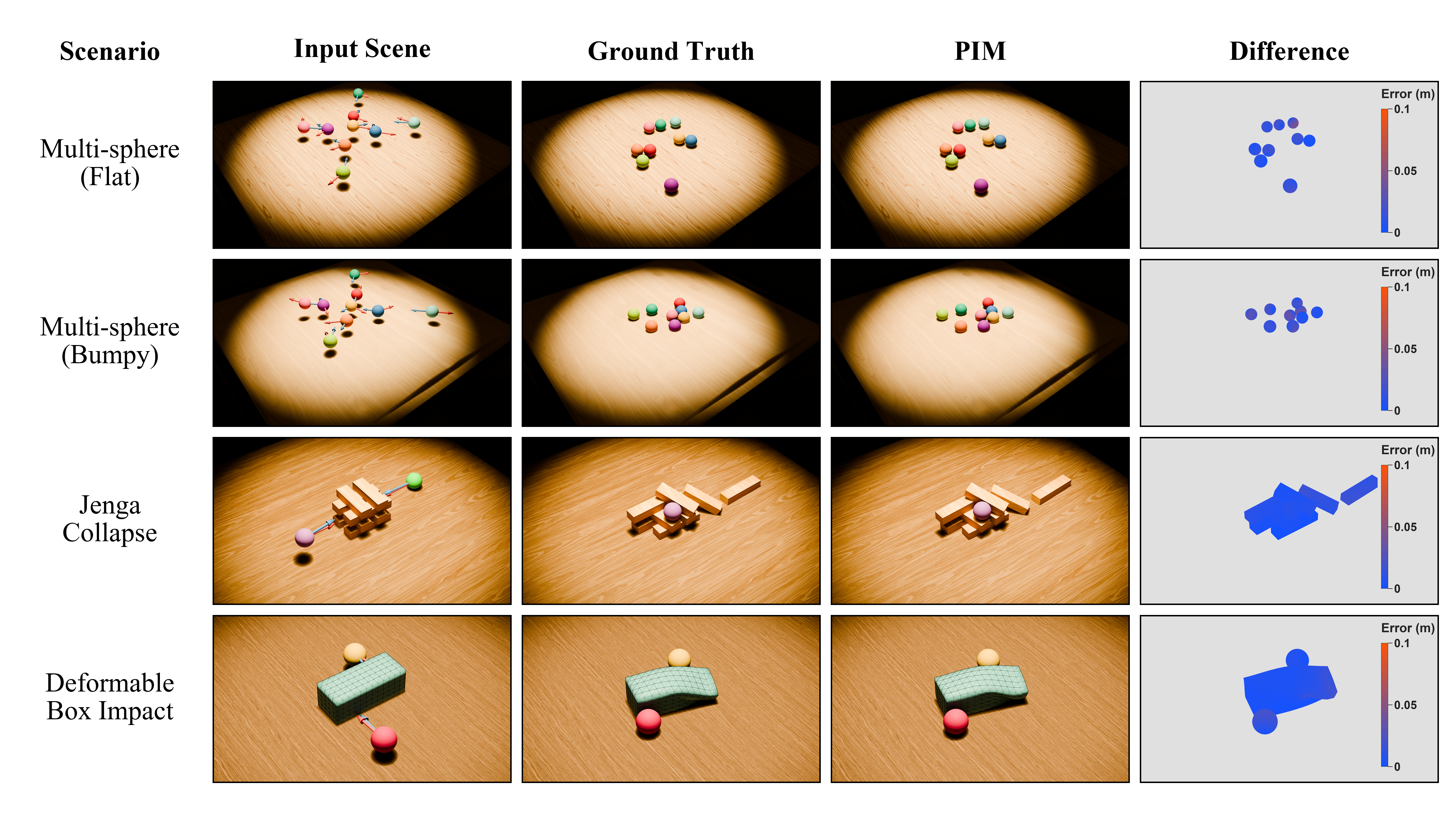}
    \caption{Qualitative results. Columns: initial scene, MuJoCo final state, PIM prediction, and vertex error. Red/blue arrows denote initial forces/velocities.}
    \label{fig:qualitative_results}
\end{figure*}

\begin{figure*}[!t]
    \centering
    \includegraphics[width=\linewidth]{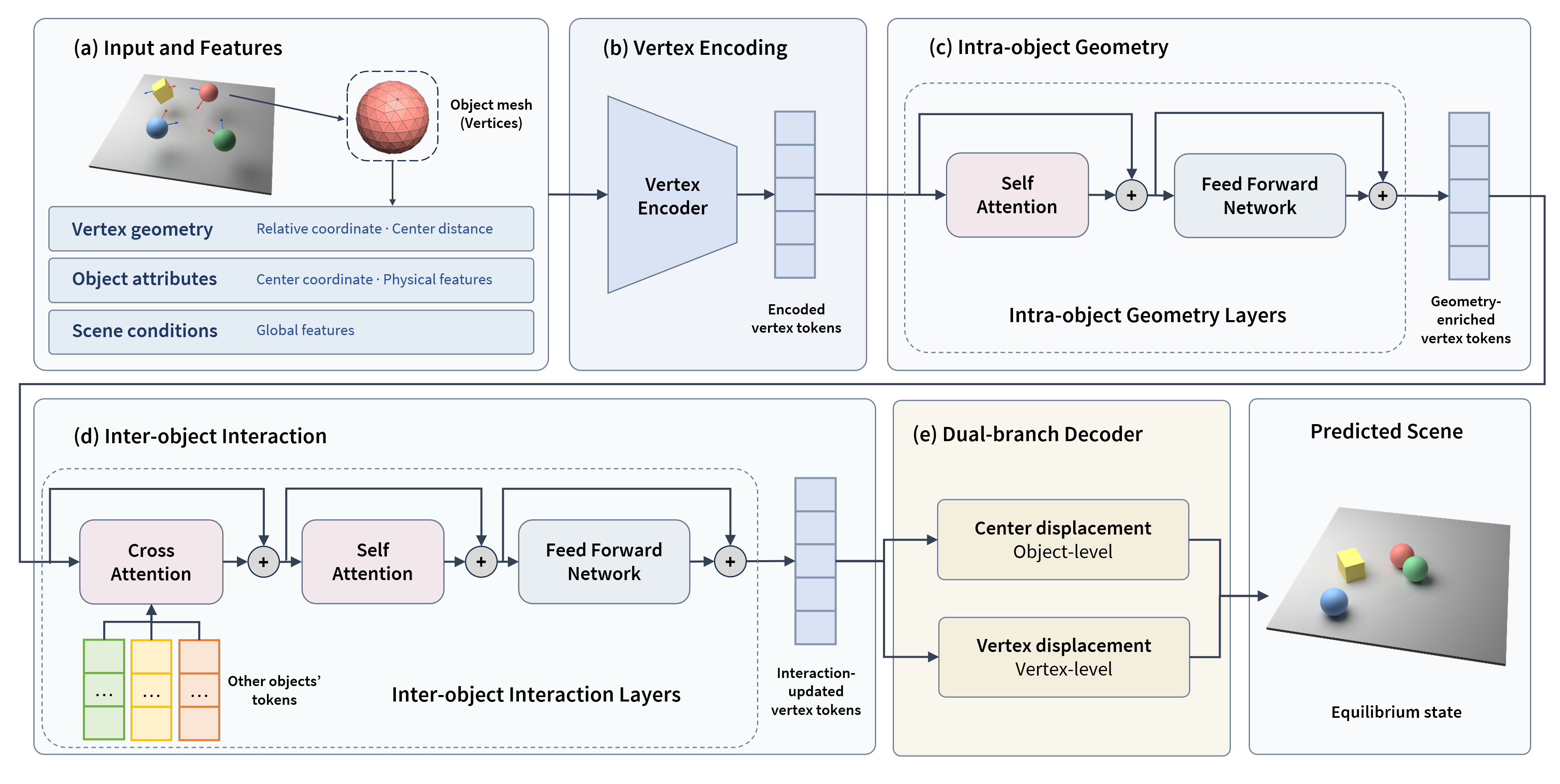}
    \caption{PIM architecture. Geometry and physical conditions become vertex tokens. Local KNN attention encodes geometry, anchor-based attention models cross-object interactions, and dual-branch decoding predicts object translation and vertex residuals.}
    \label{fig:architecture}
\end{figure*}

\section{Related work}

\paragraph{Physical intuition and learning-based dynamics prediction.}
Physical intuition has been studied through outcome prediction, object-centric reasoning, and benchmarks covering phenomena such as collision, rolling, sliding, and soft-body interaction~\cite{lerer2016physical,piloto2022intuitive,bear2021physion}. Learning-based dynamics models approximate temporal physical evolution using object-centric interaction functions and visual interaction networks~\cite{battaglia2016interactionnet,watters2017vin}, as well as particle- or graph-based message passing~\cite{li2019dpinet,sanchez2020gns,yang2022gpe}. Later work studies generalization, symmetry, contact handling, and rigid-body prediction~\cite{han2022subequivariant,allen2022gns_contact,rubanova2024sdfsim}, while HOPNet uses time-varying combinatorial complexes for autoregressive rigid-body prediction~\cite{Wei2025}. Other object-centric methods learn latent representations for future-frame, trajectory, or dynamics prediction~\cite{schmeckpeper2021opa,min2021gatsbi,nakano2023stedie,song2025ock}. PIM differs from these predominantly temporal formulations by predicting an explicit mesh-level equilibrium state from 3D geometry and physical conditions.

\paragraph{Mesh-, point-, and contact-based dynamics modeling.}
Mesh and point representations support learning on irregular geometric domains. Point Transformer models local point-set structure with self-attention~\cite{zhao2021pointtransformer}. MeshGraphNets and Boundary Graph Neural Networks learn dynamics on meshes or particle--boundary graphs~\cite{pfaff2021meshgnn,mayr2023bgnn}, while related point-sequence and mesh-surrogate methods address scene flow, point-cloud forecasting, and solid-mechanics simulation~\cite{pan2022spcmnet,deshpande2024magnet}. Contact-aware approaches study differentiable simulation of cloth, deformable bodies, and fluid--rigid interactions~\cite{newbury2024review,li2023diffcloth,huang2024diffsolver,li2024difffr}, while learned rigid-body simulators model contact through face-based interactions~\cite{allen2023fignet}. PIM complements these approaches by treating mesh vertices as prediction tokens, using local KNN attention for intra-object structure and anchor-based attention for cross-object interaction.

\paragraph{Transformer-based physical modeling.}
Transformers provide a general attention-based architecture~\cite{vaswani2017transformer}, while neural operators learn solution mappings for parametric PDEs and physical surrogate modeling~\cite{li2021fno,wang2021deeponet}. Subsequent work explores Galerkin, wavelet, orthogonal, factorized, and geometry-aware attention mechanisms~\cite{cao2021galerkin,tripura2022wno,xiao2024orthogonal,li2023factformer,wu2024transolver,alkin2024upt}, including Transformer-based simulation on irregular meshes~\cite{garnier2026ttmbs}. PIM applies this attention-based perspective to multi-object terminal prediction, combining object-specific mesh tokens with separate intra-object and inter-object interaction mechanisms.

\section{Physical intuition model}

\subsection{Problem formulation}
\label{sec:problem_formulation}

For a scene containing $M$ objects, let $\mathbf{X}_i^0\in\mathbb{R}^{n_i\times3}$ denote the initial vertices of object $i$, and let $\mathbf{Y}_i$ and $\widehat{\mathbf{Y}}_i$ denote its simulated and predicted final vertices. Each object has initial velocity $\mathbf{v}_i$, an external-force vector $\mathbf{f}_i$ applied during the initial forcing interval, density $\rho_i$, and sliding, spinning, and rolling friction $\bm{\mu}_i=[\mu_i^t,\mu_i^s,\mu_i^r]$. We normalize force and density as $\widetilde{\mathbf{f}}_i=\mathbf{f}_i/10$ and $\widetilde{\rho}_i=\rho_i/2000$, and define the object condition $\mathbf{o}_i=[\mathbf{v}_i,\widetilde{\mathbf{f}}_i,\widetilde{\rho}_i,\bm{\mu}_i]\in\mathbb{R}^{10}$. Surface friction and normalized vertical gravity form $\mathbf{g}=[\mu_{\mathrm{surf}}^t,\mu_{\mathrm{surf}}^s,\mu_{\mathrm{surf}}^r,g_z/10]\in\mathbb{R}^{4}$. PIM learns
\begin{equation}
\begin{aligned}
\mathcal{F}_\theta:
&\left\{(\mathbf{X}_i^0,\mathbf{o}_i)\right\}_{i=1}^{M}\cup\{\mathbf{g}\}
\\
&\quad
\longrightarrow
\left\{\widehat{\mathbf{Y}}_i\right\}_{i=1}^{M}.
\end{aligned}
\end{equation}

\subsection{Architecture overview}

PIM encodes geometry and physical conditions into vertex features, applies local KNN and anchor-based cross-object attention, and decodes object translation and vertex residuals (Figure~\ref{fig:architecture}).

\subsection{Input encoding}

For center $\mathbf{c}_i^0=\frac{1}{n_i}\sum_p\mathbf{x}_{i,p}^0$, define relative coordinates $\mathbf{r}_{i,p}^0=\mathbf{x}_{i,p}^0-\mathbf{c}_i^0$ and radius $s_{i,p}^0=\|\mathbf{r}_{i,p}^0\|_2$. Independent projections of these quantities, the center, and object/scene conditions have dimensions $d/2$, $d/16$, $d/8$, $d/4$, and $d/16$, respectively:
\begin{equation}
\begin{aligned}
\mathbf{h}_{i,p}=\bigl[
&
\phi_{\mathrm{rel}}(\mathbf{r}_{i,p}^0),
\phi_{\mathrm{dist}}(s_{i,p}^0),
\phi_{\mathrm{cen}}(\mathbf{c}_i^0),\\
&
\phi_{\mathrm{obj}}(\mathbf{o}_i),
\phi_{\mathrm{glb}}(\mathbf{g})
\bigr]\in\mathbb{R}^d.
\end{aligned}
\end{equation}
A fusion projection and layer normalization yield $\mathbf{H}_{i,p}^0=\operatorname{LN}(\phi_{\mathrm{fuse}}(\mathbf{h}_{i,p}))$ and the initial representation $\mathbf{H}_i^0\in\mathbb{R}^{n_i\times d}$.

\subsection{Intra- and inter-object attention}

\begin{figure}[!htb]
    \centering
    \includegraphics[width=0.5\linewidth]{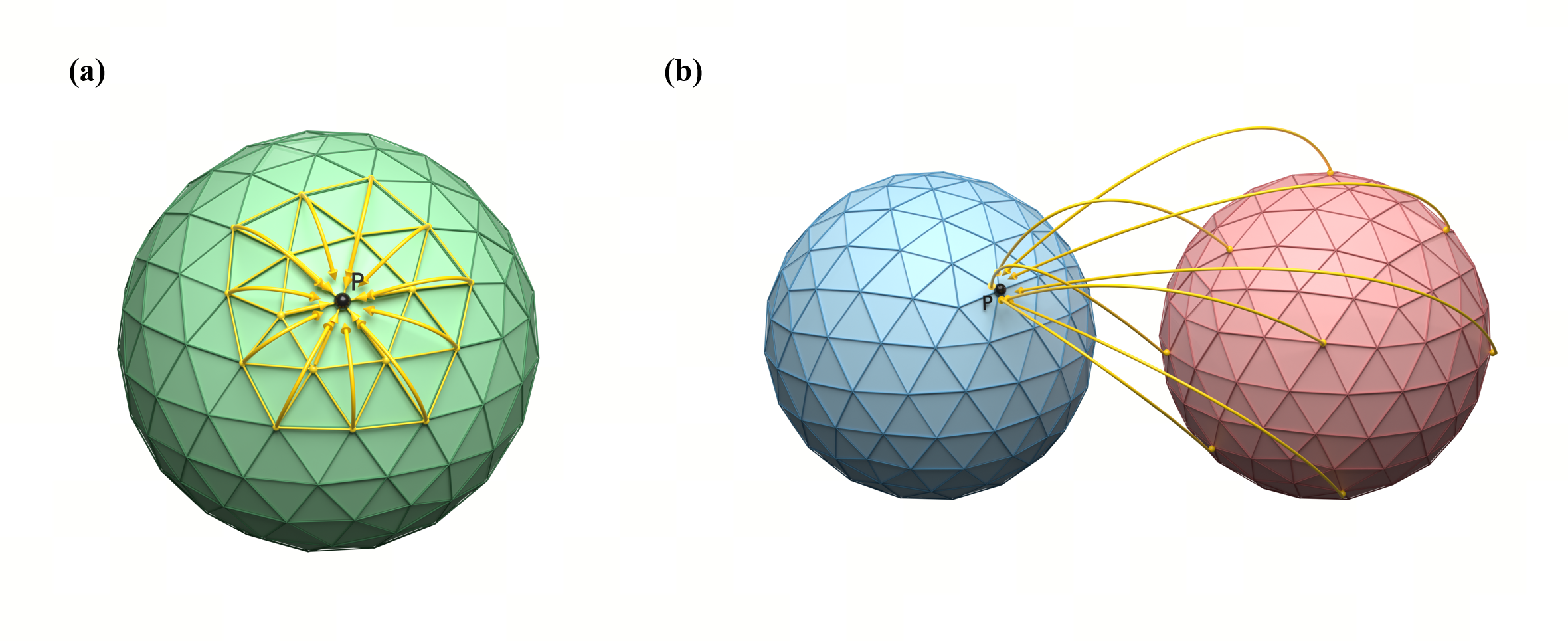}
    \caption{Local and cross-object attention. A target aggregates from (a) its object's KNN neighbors or (b) another object's anchors. Arrows point toward the target.}
    \label{fig:attention}
\end{figure}

A fixed initial-space KNN graph defines neighborhood $\mathcal{N}_i(p)$. Local attention aggregates:
\begin{equation}
\begin{aligned}
\mathbf{z}_{i,p}
&=
\sum_{q\in\mathcal{N}_i(p)}
\alpha_{p,q}\mathbf{u}_{i,q},\\
\alpha_{p,q}
&=
\operatorname{softmax}_{q\in\mathcal{N}_i(p)}
\left(
\overline{\mathbf{q}}_{i,p}^{\mathsf T}
\overline{\mathbf{k}}_{i,q}
\right),
\end{aligned}
\end{equation}
Queries and keys are L2-normalized, giving cosine logits without temperature scaling.

Each source object $j$ supplies $a$ initial-geometry anchors $\mathcal{A}_j$ through deterministic FPS, starting at vertex zero:
\begin{equation}
\begin{aligned}
\pi_{j,0}&=0,\\
\pi_{j,l}
&=
\arg\max_q
\min_{r<l}
\left\|
\mathbf{x}_{j,q}^{0}
-
\mathbf{x}_{j,\pi_{j,r}}^{0}
\right\|_2^2,
\\
&\quad l=1,\ldots,a-1.
\end{aligned}
\end{equation}
A target vertex of object $i$ attends to anchors of every $j\neq i$:
\begin{equation}
\begin{aligned}
\mathbf{z}_{i\leftarrow j,p}
&=
\sum_{q\in\mathcal{A}_j}
\beta_{p,q}^{i\leftarrow j}\mathbf{u}_{j,q},\\
\beta_{p,q}^{i\leftarrow j}
&=
\operatorname{softmax}_{q\in\mathcal{A}_j}
\left(
\overline{\mathbf{q}}_{i,p}^{\mathsf T}
\overline{\mathbf{k}}_{j,q}
\right).
\end{aligned}
\end{equation}
Messages $\mathbf{Z}_i^{\mathrm{cross}}=\frac{1}{M-1}\sum_{j\neq i}\mathbf{Z}_{i\leftarrow j}$ enter a residual update followed by local attention and feed-forward refinement. Figure~\ref{fig:attention} illustrates both attention patterns.

\subsection{Dual-branch decoder}

Mean-pooled features predict translation $\Delta\mathbf{c}_i$; the relative branch predicts vertex displacements with their object-wise mean removed:
\begin{equation}
\begin{aligned}
\Delta\mathbf{c}_i
&=
\operatorname{MLP}_c\!\left(\frac{1}{n_i}\sum_p\mathbf{H}_{i,p}^{\mathrm{int}}\right),\\
\widetilde{\Delta\mathbf{r}}_{i,p}
&=
\operatorname{MLP}_r(\mathbf{H}_{i,p}^{\mathrm{int}})\\
&\quad-
\frac{1}{n_i}\sum_q\operatorname{MLP}_r(\mathbf{H}_{i,q}^{\mathrm{int}}).
\end{aligned}
\end{equation}
The final vertex is
\begin{equation}
\widehat{\mathbf{y}}_{i,p}
=
\mathbf{x}_{i,p}^{0}
+\Delta\mathbf{c}_i
+\widetilde{\Delta\mathbf{r}}_{i,p}.
\end{equation}
Zero-initialized output layers initially reproduce the input mesh.

\subsection{Loss function}

Let $\widehat{\mathbf{c}}_{b,i}$ and $\mathbf{c}_{b,i}$ be predicted and target vertex means, and define pairwise distances
\begin{equation}
\begin{aligned}
\ell^{\mathrm{pred}}_{b,i,p,q}
&=
\|\widehat{\mathbf{y}}_{b,i,p}-\widehat{\mathbf{y}}_{b,i,q}\|_2,\\
\ell^{\mathrm{target}}_{b,i,p,q}
&=
\|\mathbf{y}_{b,i,p}-\mathbf{y}_{b,i,q}\|_2.
\end{aligned}
\end{equation}
For batch size $B$, the three supervised losses are
\begin{align}
\mathcal{L}_{\mathrm{coord}}
&=
\frac{1}{3B\sum_i n_i}
\sum_{b,i,p}
\left\|\widehat{\mathbf{y}}_{b,i,p}-\mathbf{y}_{b,i,p}\right\|_1,\\
\mathcal{L}_{\mathrm{cen}}
&=
\frac{1}{BM}\sum_{b,i}
\left\|\widehat{\mathbf{c}}_{b,i}-\mathbf{c}_{b,i}\right\|_2^2,\\
\mathcal{L}_{\mathrm{edge}}
&=
\frac{1}{B\sum_i n_i k_i}
\sum_{b,i,p}\sum_{q\in\mathcal{N}_i(p)}
\left|
\ell^{\mathrm{pred}}_{b,i,p,q}
-
\ell^{\mathrm{target}}_{b,i,p,q}
\right|.
\end{align}
Edge supervision and evaluation use a fixed initial-space 32-neighbor graph, with $k_i=\min(32,n_i-1)$. Final-mesh target lengths permit deformation. Training minimizes
\begin{equation}
\mathcal{L}
=
\mathcal{L}_{\mathrm{coord}}
+\lambda_{\mathrm{cen}}\mathcal{L}_{\mathrm{cen}}
+\lambda_{\mathrm{edge}}\mathcal{L}_{\mathrm{edge}}.
\label{eq:training_objective}
\end{equation}
Coarse training uses $(\lambda_{\mathrm{cen}},\lambda_{\mathrm{edge}})=(1,5)$, ramping the edge weight from zero over 10 epochs. Fine training uses $(0.1,20)$.

\section{Experiments}

\begin{table*}[!t]
\centering
\caption{Direct prediction and fixed-checkpoint Jenga/Deformable generalization (meters; $\downarrow$). Indented rows evaluate the named object under the preceding OOD condition; other rows are scene-level. Jenga and Deformable are ID references. Bold indicates the lowest vertex and centroid errors. Edge errors are geometric diagnostics; $^\dagger$ denotes rigid outputs whose edge lengths are constrained by construction.}
\label{tab:baseline_results}
\label{tab:jenga_generalization_results}
\small
\setlength{\tabcolsep}{2pt}
\resizebox{\linewidth}{!}{%
\begin{tabular}{lrrrrrrrrr}
\toprule
& \multicolumn{3}{c}{PIM}
& \multicolumn{3}{c}{Vertex Transformer}
& \multicolumn{3}{c}{Rigid-Pose MLP} \\
\cmidrule(lr){2-4}\cmidrule(lr){5-7}\cmidrule(lr){8-10}
Scenario / condition
& $\varepsilon_{\mathrm{vtx}}$ & $\varepsilon_{\mathrm{cen}}$ & $\varepsilon_{\mathrm{edge}}$
& $\varepsilon_{\mathrm{vtx}}$ & $\varepsilon_{\mathrm{cen}}$ & $\varepsilon_{\mathrm{edge}}$
& $\varepsilon_{\mathrm{vtx}}$ & $\varepsilon_{\mathrm{cen}}$ & $\varepsilon_{\mathrm{edge}}$ \\
\midrule
Flat & \textbf{0.025695} & \textbf{0.020983} & $1.182\times10^{-5}$ & 0.035674 & 0.033369 & $3.896\times10^{-5}$ & 0.036876 & 0.035471 & $7.732\times10^{-6}\,{}^\dagger$ \\
Bumpy & \textbf{0.027548} & \textbf{0.025536} & $2.533\times10^{-5}$ & 0.033929 & 0.032083 & $3.455\times10^{-5}$ & 0.035064 & 0.033051 & $7.690\times10^{-6}\,{}^\dagger$ \\
Jenga & \textbf{0.020358} & \textbf{0.019961} & $8.217\times10^{-5}$ & 0.029537 & 0.028590 & $5.751\times10^{-5}$ & 0.029298 & 0.029221 & $1.071\times10^{-5}\,{}^\dagger$ \\
Deformable & \textbf{0.019305} & \textbf{0.013211} & $3.805\times10^{-4}$ & 0.044912 & 0.021323 & $3.680\times10^{-4}$ & 0.032454 & 0.019386 & $7.889\times10^{-4}\,{}^\dagger$ \\
\midrule
\multicolumn{10}{l}{\textit{Jenga generalization}} \\
ID-U & \textbf{0.021130} & \textbf{0.020458} & $8.203\times10^{-5}$ & 0.029082 & 0.027877 & $5.717\times10^{-5}$ & 0.029089 & 0.029782 & $1.070\times10^{-5}\,{}^\dagger$ \\
Position OOD & \textbf{0.025523} & \textbf{0.023533} & $8.793\times10^{-5}$ & 0.035871 & 0.035286 & $6.358\times10^{-5}$ & 0.033835 & 0.034743 & $\mathbf{1.070\times10^{-5}}\,{}^\dagger$ \\
Velocity OOD & \textbf{0.033473} & \textbf{0.031966} & $9.128\times10^{-5}$ & 0.040956 & 0.040407 & $6.464\times10^{-5}$ & 0.042400 & 0.044198 & $1.068\times10^{-5}\,{}^\dagger$ \\
Force OOD & \textbf{0.024587} & \textbf{0.023186} & $8.501\times10^{-5}$ & 0.034687 & 0.032747 & $6.252\times10^{-5}$ & 0.034308 & 0.034856 & $1.071\times10^{-5}\,{}^\dagger$ \\
Shape OOD & \textbf{0.036957} & \textbf{0.032656} & $1.182\times10^{-4}$ & 0.041459 & 0.036174 & $3.737\times10^{-4}$ & 0.045558 & 0.046572 & $1.055\times10^{-5}\,{}^\dagger$ \\
\quad Replaced cube & \textbf{0.074302} & \textbf{0.064307} & $8.149\times10^{-4}$ & 0.128112 & 0.115361 & $6.499\times10^{-3}$ & 0.193574 & 0.188679 & $1.096\times10^{-5}\,{}^\dagger$ \\

\midrule
\multicolumn{10}{l}{\textit{Deformable generalization}} \\
Position OOD & \textbf{0.028668} & \textbf{0.020162} & $4.707\times10^{-4}$ & 0.054941 & 0.029873 & $4.025\times10^{-4}$ & 0.039837 & 0.027239 & $6.787\times10^{-4}\,{}^\dagger$ \\
\quad Soft box & \textbf{0.009342} & \textbf{0.005582} & $1.792\times10^{-3}$ & 0.012488 & 0.007856 & $1.965\times10^{-3}$ & 0.016568 & 0.009090 & $3.619\times10^{-3}\,{}^\dagger$ \\
Velocity OOD & \textbf{0.029399} & \textbf{0.021192} & $4.869\times10^{-4}$ & 0.050068 & 0.028628 & $3.832\times10^{-4}$ & 0.039981 & 0.027875 & $6.349\times10^{-4}\,{}^\dagger$ \\
\quad Soft box & \textbf{0.011341} & \textbf{0.007112} & $1.771\times10^{-3}$ & 0.016047 & 0.009470 & $1.874\times10^{-3}$ & 0.017782 & 0.011266 & $3.383\times10^{-3}\,{}^\dagger$ \\
Force OOD & \textbf{0.022031} & \textbf{0.015752} & $3.801\times10^{-4}$ & 0.046189 & 0.022650 & $3.544\times10^{-4}$ & 0.029953 & 0.020658 & $6.649\times10^{-4}\,{}^\dagger$ \\
\quad Soft box & \textbf{0.008278} & \textbf{0.004661} & $1.624\times10^{-3}$ & 0.011321 & 0.006612 & $1.758\times10^{-3}$ & 0.015274 & 0.008170 & $3.544\times10^{-3}\,{}^\dagger$ \\
\bottomrule
\end{tabular}
}
\end{table*}

\subsection{Experimental setup}

\paragraph{Datasets.}
We train one PIM per MuJoCo~\cite{todorov2012mujoco} scenario: nine spheres interacting on flat (Flat) or irregular (Bumpy) surfaces, impact-induced collapse of an eight-block tower (Jenga), and two rigid spheres impacting a deformable box (Deformable).

Flat and Bumpy each contain nine objects and 5,778 fine vertices per scene, with 30,000/1,000/1,000 training/validation/test scenes. Jenga has ten objects and 2,116 vertices with 40,000/2,000/2,000 scenes; Deformable has three objects and 1,577 vertices with 5,000/200/200 scenes. Splits use disjoint random-number streams and split-specific seeds; object ordering and mesh topology are fixed within each scenario. Samples are filtered for multiple impacts on Flat/Bumpy, impact-induced collapse on Jenga, or rigid--soft contact with measurable deformation on Deformable.

MuJoCo uses a $0.002\,\mathrm{s}$ time step and applies external forces for the first $0.05\,\mathrm{s}$. Target meshes are taken at termination: at most $10\,\mathrm{s}$ for Flat/Bumpy/Jenga or $8\,\mathrm{s}$ for Deformable, with early termination when the maximum absolute generalized velocity stays below $10^{-3}$ for $0.2\,\mathrm{s}$. The controlled Flat comparison instead uses fixed $2\,\mathrm{s}$ targets.

\begin{figure*}[!t]
    \centering
    \includegraphics[width=\textwidth]{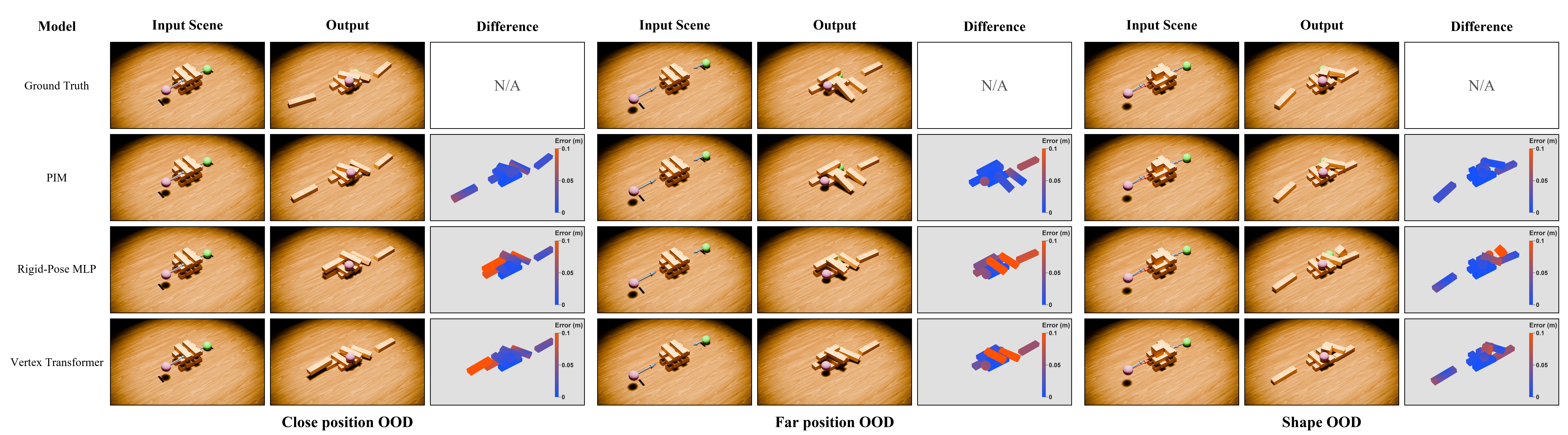}
    \caption{Matched-protocol Jenga OOD: close position, far position and shape (left to right). Rows: ground truth and three methods; columns: input, final state and vertex error. Black segments mark close/far ID boundaries; red/blue arrows denote initial forces/velocities.}
    \label{fig:jenga_ood_qualitative}
\end{figure*}

\begin{figure*}[!t]
    \centering
    \includegraphics[width=\textwidth,height=0.55\textheight,keepaspectratio]{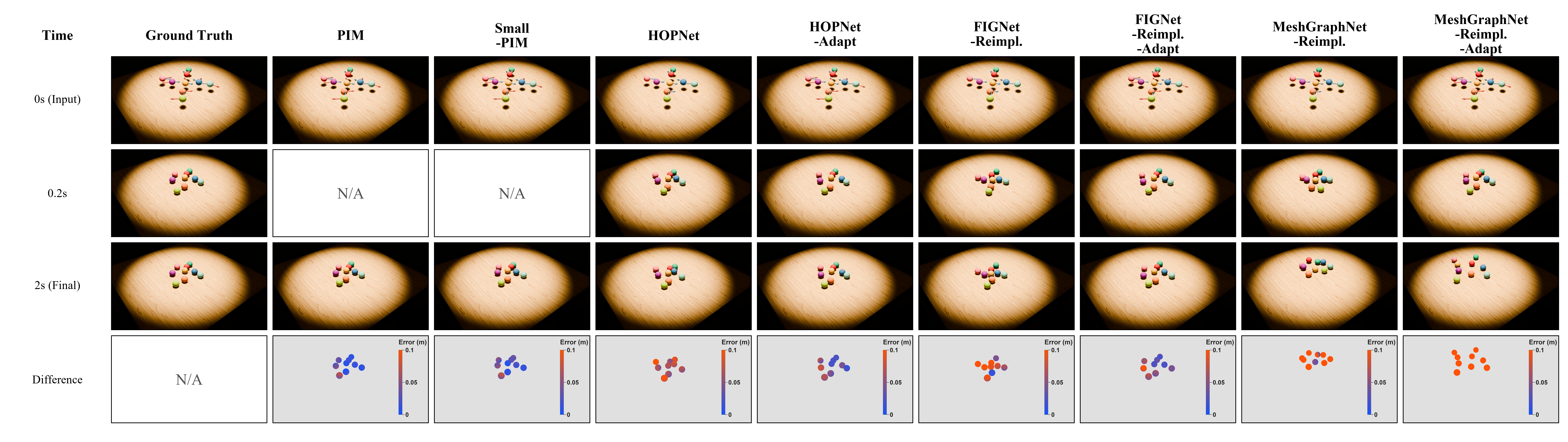}
    \caption{One held-out controlled Flat scene. Rows: ground truth, PIM, Small-PIM, and six rollout baselines. Columns: input ($t=0$), intermediate state ($0.2\,\mathrm{s}$), equilibrium ($2\,\mathrm{s}$), and object-wise terminal errors. Direct predictors output no intermediate state. Reimpl.\ denotes our PyTorch reimplementation.}
    \label{fig:rollout_comparison}
\end{figure*}

\begin{figure}[!htb]
    \centering
    \includegraphics[width=0.9\linewidth]{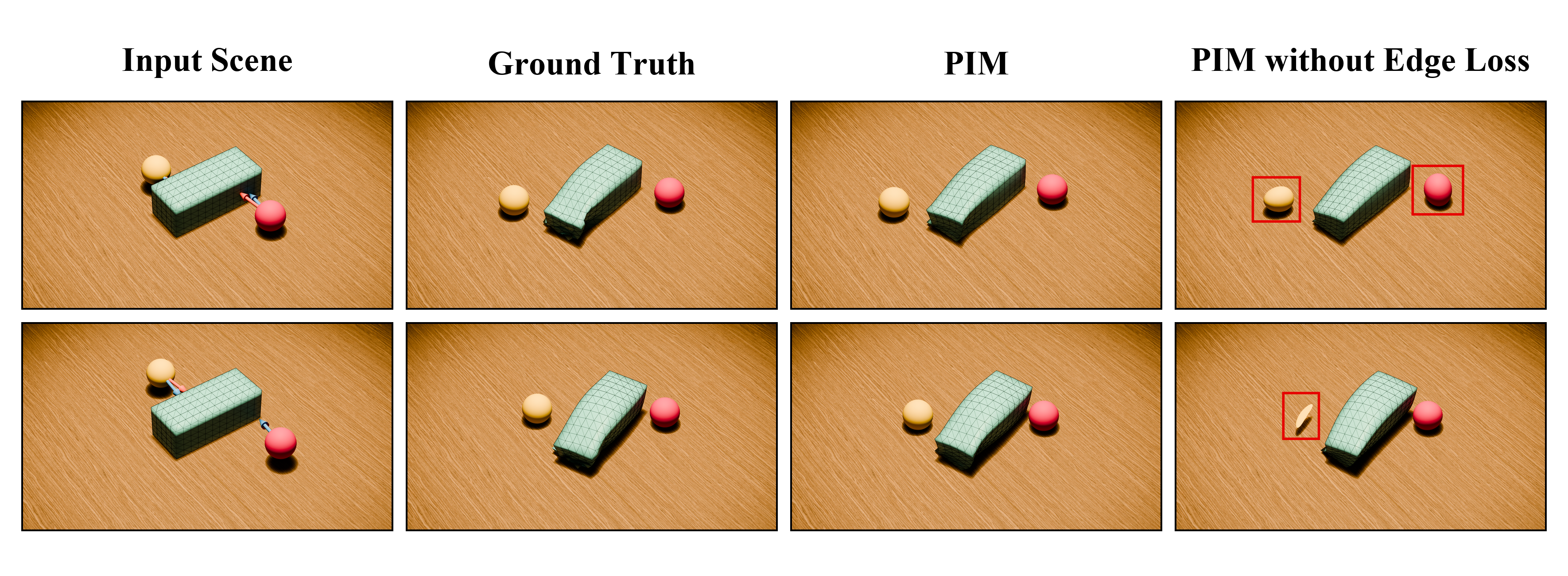}
    \caption{Deformable edge-loss ablation. Columns: input, ground truth, full PIM, no edge loss. Red boxes highlight distortions without edge supervision. Results over all 200 test scenes: Table~\ref{tab:ablation_results}.}
    \label{fig:no_edge_comparison}
\end{figure}

\begin{figure*}[!t]
    \centering
    \includegraphics[width=\textwidth,height=0.45\textheight,keepaspectratio]{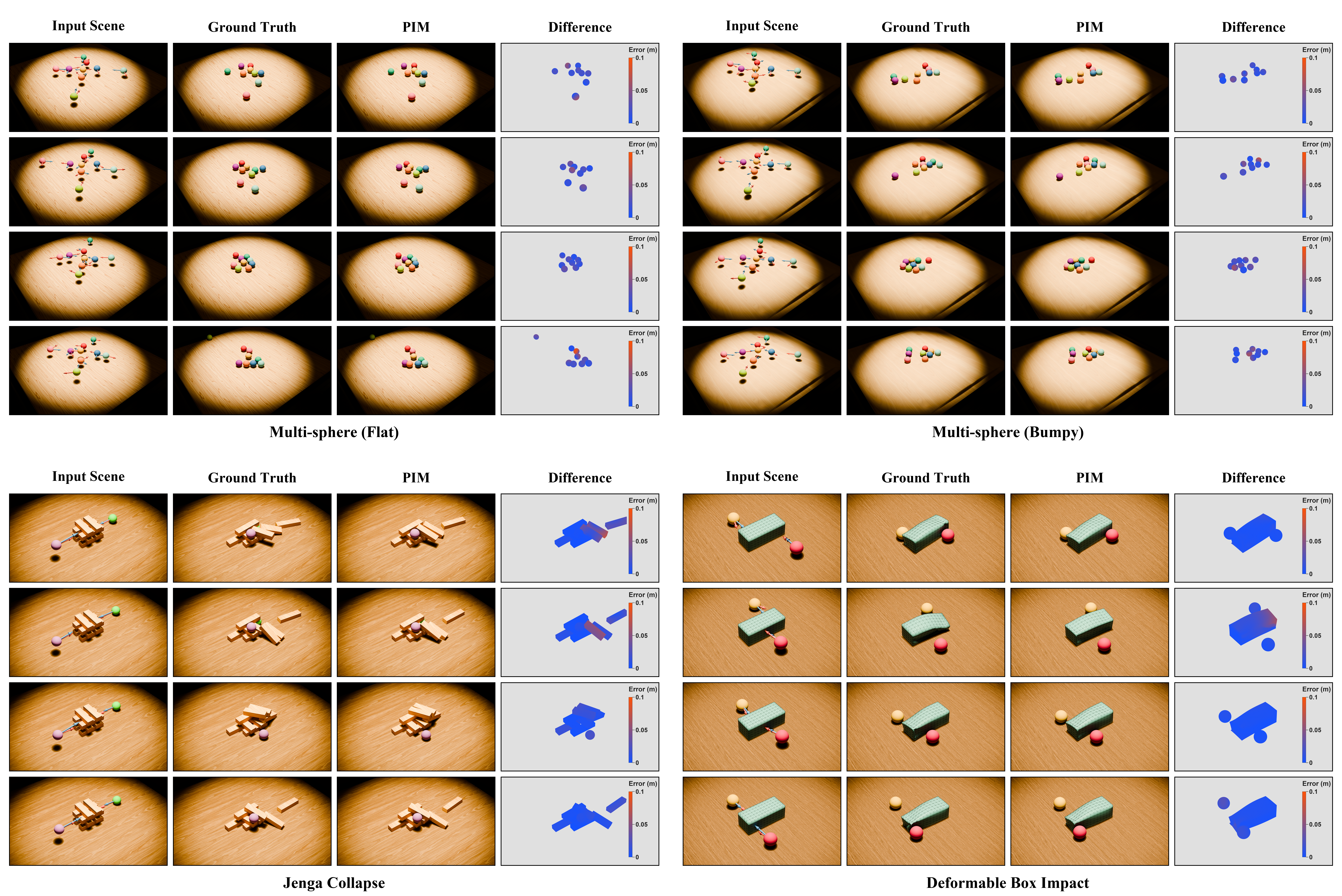}
    \caption{Four test scenes per scenario. Rows: scenes; columns: input, MuJoCo final state, PIM, vertex error. Red/blue arrows denote initial forces/velocities.}
    \label{fig:qualitative_all}
\end{figure*}

\paragraph{Model and training.}
Flat/Bumpy use width 512, eight heads, two local layers and two interaction layers; Jenga/Deformable use width 768, six heads, four local layers and two interaction layers. Defaults are $k=a=32$ and FFN expansion four. Coarse training uses stride-four vertex subsampling; its validation-selected checkpoint initializes fine training, which recomputes features and neighborhoods on the full mesh. Only the fine model is used at inference.

PIM uses AdamW~\cite{loshchilov2019adamw} with linear warm-up and cosine decay. Coarse training runs for 80 epochs on Flat/Bumpy and 100 on Jenga/Deformable, starting at $3\times10^{-4}$. Fine training runs for 40 epochs, starting at $1.2\times10^{-4}$, except Bumpy uses 20 epochs starting at $2\times10^{-5}$. Both stages select checkpoints by validation coordinate L1 error. Full-dataset PIM training uses four RTX A6000 GPUs.

\paragraph{Baselines.}
To our knowledge, no publicly available model directly supports our multi-object equilibrium-mesh prediction setting. We therefore construct two direct equilibrium-state baselines: Rigid-Pose MLP predicts object translation and rotation from pooled geometry/physics descriptors, while Vertex Transformer predicts vertex displacements using local attention and a Transformer over pooled object tokens. Both use the same fine meshes, splits and supervised loss as PIM, with the direct-baseline weights given above; they train for 100 epochs using AdamW at an initial $10^{-4}$ and validation-L1 checkpoint selection.

A controlled Flat comparison evaluates PIM and Small-PIM against HOPNet~\cite{Wei2025} and our PyTorch reimplementations of FIGNet~\cite{allen2023fignet,Yi_FIGNet_PyTorch_implementation_2024} and MeshGraphNets~\cite{pfaff2021meshgnn}; \emph{-Adapt} variants also receive the initial external force. We report mean vertex and centroid errors for the controlled Flat comparison. The direct baselines are additionally evaluated on Jenga and Deformable generalization sets without fine-tuning.

\begin{table*}[!t]
\centering
\caption{Controlled Flat at $t=2\,\mathrm{s}$. Errors are in meters ($\downarrow$); parameters are in millions (rounded). FIGNet/MGN denote our FIGNet/MeshGraphNet reimplementations; -A denotes Adapt. $^\dagger$ Triangle-edge errors reflect rigid outputs, per-step rigid projection, or terminal rigid-pose transfer.}
\label{tab:rollout_baseline_results}
\small
\setlength{\tabcolsep}{2pt}
\begin{tabular}{@{}lrrrr@{}}
\toprule
Method & \shortstack{Params.\\(M)} & $\varepsilon_{\mathrm{vtx}}$ & $\varepsilon_{\mathrm{cen}}$ & $\varepsilon_{\mathrm{tri}}$ \\
\midrule
PIM & 19.7556 & \textbf{0.030731} & \textbf{0.027970} & $9.03\times10^{-6}$ \\
Small-PIM & 5.0133 & 0.031991 & 0.029233 & $9.79\times10^{-6}$ \\
HOPNet & 0.7071 & 0.049890 & 0.048495 & $6.44\times10^{-9}\,{}^\dagger$ \\
HOPNet-A & 0.7075 & 0.038674 & 0.037202 & $6.44\times10^{-9}\,{}^\dagger$ \\
FIGNet & 5.5605 & 0.079846 & 0.078385 & $7.13\times10^{-9}\,{}^\dagger$ \\
FIGNet-A & 5.5612 & 0.043898 & 0.041480 & $7.17\times10^{-9}\,{}^\dagger$ \\
MGN & 2.8787 & 0.197970 & 0.194698 & $7.90\times10^{-9}\,{}^\dagger$ \\
MGN-A & 2.8791 & 0.219304 & 0.215845 & $8.42\times10^{-9}\,{}^\dagger$ \\
\bottomrule
\end{tabular}
\end{table*}

\begin{table*}[!t]
\centering
\caption{Controlled ablations on each scenario's test split. Errors are in meters ($\downarrow$).}
\label{tab:ablation_results}
\small
\setlength{\tabcolsep}{1pt}
\begin{tabular}{@{}llrrr@{}}
\toprule
Scenario & Method & $\varepsilon_{\mathrm{vtx}}$ & $\varepsilon_{\mathrm{cen}}$ & $\varepsilon_{\mathrm{edge}}$ \\
\midrule
Flat & Full & \textbf{0.025695} & \textbf{0.020983} & $1.182\times10^{-5}$ \\
& Direct decoder & 0.031047 & 0.029016 & $4.739\times10^{-5}$ \\
& No cross attention & 0.060412 & 0.058884 & $1.181\times10^{-5}$ \\
\addlinespace[2pt]
Bumpy & Full & \textbf{0.027548} & \textbf{0.025536} & $2.533\times10^{-5}$ \\
& Direct decoder & 0.028434 & 0.026067 & $8.041\times10^{-5}$ \\
& No cross attention & 0.054619 & 0.053423 & $1.166\times10^{-5}$ \\
\addlinespace[2pt]
Jenga & Full & \textbf{0.020358} & \textbf{0.019961} & $8.217\times10^{-5}$ \\
& Direct decoder & 0.021999 & 0.020817 & $5.764\times10^{-5}$ \\
& No cross attention & 0.044889 & 0.060364 & $4.419\times10^{-5}$ \\
\addlinespace[2pt]
Deformable & Full & \textbf{0.019305} & \textbf{0.013211} & $3.805\times10^{-4}$ \\
& Direct decoder & 0.019547 & 0.013387 & $3.811\times10^{-4}$ \\
& No cross attention & 0.036225 & 0.027690 & $5.698\times10^{-4}$ \\
& No edge loss & 0.020123 & 0.013961 & $1.245\times10^{-3}$ \\
\bottomrule
\end{tabular}
\end{table*}

\begin{table*}[!t]
\centering
\caption{Runtime. Top: single-scene latency and matched parallel throughput across scenarios. Bottom: controlled Flat equilibrium-state prediction at $2\,\mathrm{s}$.}
\label{tab:runtime_results}
\setlength{\tabcolsep}{1.7pt}
\renewcommand{\arraystretch}{0.92}

\begin{tabular}{lrrrrrr}
\toprule
&
\multicolumn{3}{c}{\textbf{Single-scene latency}} &
\multicolumn{3}{c}{\textbf{Matched parallel throughput}} \\
\cmidrule(lr){2-4}\cmidrule(lr){5-7}
Scenario
& PIM (ms) & MuJoCo (ms) & Speedup
& PIM (samples/s) & MuJoCo (samples/s) & Speedup \\
\midrule
Flat
& 67.400 & 542.579 & $8.05\times$
& 16.779 & 11.347 & $1.48\times$ \\
Bumpy
& 68.110 & 5,083.775 & $74.64\times$
& 16.868 & 1.490 & $11.32\times$ \\
Jenga
& 66.383 & 1,686.671 & $25.41\times$
& 44.318 & 4.013 & $11.04\times$ \\
Deformable
& 27.352 & 265,110.559 & $9{,}692.51\times$
& 64.102 & 0.0202 & $3{,}173.79\times$ \\
\bottomrule
\end{tabular}

\vspace{0.35em}

\begingroup
\setlength{\tabcolsep}{5.3pt}
\begin{tabular}{lrrr}
\toprule
\multicolumn{4}{c}{\textbf{Controlled Flat terminal timing at $2\,\mathrm{s}$}} \\
\cmidrule(lr){1-4}
Method & Calls
& \shortstack{Latency$B/W{=}1$ (s)}
& \shortstack{Throughput$B/W{=}8$ (samples/s)} \\
\midrule
PIM & 1 & \textbf{0.0697} & 25.9775 \\
Small-PIM & 1 & 0.0754 & \textbf{41.0214} \\
HOPNet & 249 & 461.6632 & 0.008279 \\
HOPNet-Adapt & 249 & 406.7234 & 0.008895 \\
FIGNet-Reimpl. & 250 & 8.3254 & 0.249863 \\
FIGNet-Reimpl.-Adapt & 250 & 9.6124 & 0.243478 \\
MeshGraphNet-Reimpl. & 250 & 3.7183 & 0.792329 \\
MeshGraphNet-Reimpl.-Adapt & 250 & 3.5632 & 0.777380 \\
\bottomrule
\end{tabular}
\endgroup
\end{table*}

\paragraph{Evaluation metrics.}
For one scene, omitting batch indices in $\ell$, vertex and centroid errors are
\begin{equation}
\begin{aligned}
\varepsilon_{\mathrm{vtx}}
&=\frac{1}{\sum_i n_i}\sum_{i,p}
\|\widehat{\mathbf{y}}_{i,p}-\mathbf{y}_{i,p}\|_2,\\
\varepsilon_{\mathrm{cen}}
&=\frac{1}{M}\sum_i
\|\widehat{\mathbf{c}}_i-\mathbf{c}_i\|_2.
\end{aligned}
\label{eq:placement_errors}
\end{equation}
Using the pairwise distances defined above, edge errors are
\begin{equation}
\begin{aligned}
\varepsilon_{\mathrm{edge}}
&=\frac{\sum_{i,p}\sum_{q\in\mathcal{N}_i(p)}
|\ell^{\mathrm{pred}}_{i,p,q}-\ell^{\mathrm{target}}_{i,p,q}|}
{\sum_i n_i k_i},\\
\varepsilon_{\mathrm{tri}}
&=\frac{1}{M}\sum_i\frac{1}{|\mathcal{E}_i^{\mathrm{tri}}|}
\sum_{(p,q)\in\mathcal{E}_i^{\mathrm{tri}}}
|\ell^{\mathrm{pred}}_{i,p,q}-\ell^{\mathrm{target}}_{i,p,q}|.
\end{aligned}
\label{eq:edge_errors}
\end{equation}
Here $\mathcal{E}_i^{\mathrm{tri}}$ contains unique triangle edges of the shared 642-vertex evaluation mesh for controlled rollout comparisons. Metrics are averaged over test scenes in meters. Vertex/KNN errors weight vertices/neighbor pairs; centroid/triangle errors weight objects equally. Edge errors are rigid-motion invariant and diagnostic; rigid outputs, per-step projection, or terminal rigid fitting constrain rollout baselines' triangle-edge errors, which are excluded from placement ranking.

\subsection{Main results}

PIM has the lowest vertex and centroid errors in all four scenarios (Table~\ref{tab:baseline_results}). Rigid-Pose MLP serves as a rigid-output reference on Deformable. Figures~\ref{fig:qualitative_results} and~\ref{fig:qualitative_all} show representative and randomly sampled predictions.

\paragraph{Generalization.}
We evaluate fixed validation-selected checkpoints without fine-tuning on position, velocity, and force shifts, keeping other inputs and settings unchanged. Each set contains 2,000 Jenga or 200 Deformable scenes, equally split between lower/upper branches. Jenga uses 90\% matched and 10\% outcome-unfiltered scenes; ID-U contains 2,000 ID scenes without impact/collapse filtering. All Deformable scenes retain rigid--soft contact and at least $0.002\,\mathrm{m}$ maximum observed soft-vertex displacement.

Lower/upper OOD intervals lie below/above the corresponding ID interval in absolute magnitude, using $x$ components in Jenga and $y$ components in Deformable.

PIM achieves the lowest scene-level vertex/centroid errors for all three shifts in both scenarios (Table~\ref{tab:baseline_results}). Across models, Jenga ID-U errors differ from matched ID by at most $3.8\%$/$2.5\%$, respectively. Jenga additionally includes 2,000 matched shape-OOD scenes replacing top block J with an equal-volume cube while retaining its vertex count and mesh connectivity; PIM reduces scene-level vertex/centroid errors by $10.9\%$/$9.7\%$ against the strongest direct baseline.

Object-specific rows isolate the replaced cube (104 of 2,116 vertices) and soft box (293 of 1,577). PIM has the lowest vertex/centroid errors on both objects. Relative to Vertex Transformer, cube vertex/centroid errors decrease by $42.0\%$/$44.3\%$, while soft-box vertex errors decrease by $25.2\%$, $29.3\%$, and $26.9\%$ under position, velocity, and force OOD, respectively. Vertex Transformer yields lower Deformable scene-level edge errors; PIM yields lower soft-box edge errors under all three shifts. Figure~\ref{fig:jenga_ood_qualitative} shows Jenga OOD predictions.

\paragraph{Controlled Flat equilibrium-state comparison.}

All methods share the 1,000/100/100 Flat split, $2\,\mathrm{s}$ targets and 642-vertex evaluation meshes. Trajectories contain 250 transitions at $0.008\,\mathrm{s}$ intervals; HOPNet takes 249 updates after its input history, and FIGNet/MeshGraphNet take 250 from $t=0$. Representations, objectives, supervised transition counts and training budgets remain method-specific.

Small-PIM uses width 256, four heads, three local layers and FFN expansion three, retaining PIM's representation, interaction layers, decoder, losses and 80/40-epoch schedule. FIGNet/MeshGraphNet use 64 vertices per sphere with rigid-pose transfer to the common mesh; FIGNet projects after every update, while MeshGraphNet fits only the terminal pose without feedback.

PIM and Small-PIM reduce vertex error by $20.5\%$ and $17.3\%$ versus HOPNet-Adapt (Table~\ref{tab:rollout_baseline_results}). Figure~\ref{fig:rollout_comparison} compares their predictions.

\subsection{Ablations}

\paragraph{No cross attention.}
We remove only cross-object updates, retaining local/FFN blocks and residual connections. Vertex error rises by $87.6\%$--$135.1\%$ and centroid error by $109.2\%$--$202.4\%$ (Table~\ref{tab:ablation_results}), demonstrating the need for inter-object communication.

\paragraph{Direct decoder.}
We replace both decoder branches with a zero-initialized per-vertex MLP. Vertex error rises by $20.8\%$, $3.2\%$, $8.1\%$, and $1.3\%$ on Flat, Bumpy, Jenga, and Deformable, respectively, and centroid error increases throughout. Separating translation and residual motion therefore consistently improves placement in these single-run comparisons.

\paragraph{No edge loss.}
We remove edge supervision in both stages and initialize fine training from the matching no-edge coarse checkpoint. On Deformable, vertex/centroid errors rise by $4.2\%$/$5.7\%$, while KNN edge-length error increases to $3.27\times$ its original value, exposing local shape degradation masked by placement metrics. Figure~\ref{fig:no_edge_comparison} shows the resulting distortions.

\subsection{Runtime}
Table~\ref{tab:runtime_results} measures PIM forward passes on an RTX A6000 and MuJoCo on two Xeon Gold 6330 CPUs, including initialization, rollout and vertex extraction for MuJoCo. Controlled Flat uses the same GPU, including transfers, dynamic graph construction, inference, state updates and rigid fitting. $B/W$ denotes PIM batch size or rollout workers, with one for latency and eight for throughput.

On controlled Flat, PIM/Small-PIM achieve $53.4\times$/$49.3\times$ lower latency and $32.8\times$/$51.8\times$ higher throughput than MeshGraphNet-Reimpl., the fastest rollout implementation without additional force input. These results demonstrate the computational time advantage of our method over the evaluated rollout baselines.

\section{Conclusion}

PIM directly predicts equilibrium meshes from initial geometry and physical conditions using local and cross-object attention with dual-branch decoding. Across four MuJoCo scenarios, it achieves lower terminal-state errors than the evaluated baselines with low latency. Ablations support cross-object communication and edge supervision.

Evaluation uses scenario-specific models, fixed object correspondences/topologies. Generalization to new object counts, topologies, and larger deformations remains open. Future work includes explicit environments and joint multi-scenario training. 

\bibliographystyle{unsrt}
\bibliography{references}

\clearpage

\setcounter{section}{0}
\setcounter{equation}{0}
\setcounter{figure}{0}
\setcounter{table}{0}

\renewcommand{\thesection}{S\arabic{section}}
\renewcommand{\theequation}{S\arabic{equation}}
\renewcommand{\thefigure}{S\arabic{figure}}
\renewcommand{\thetable}{S\arabic{table}}

\renewcommand{\theHsection}{supp.\arabic{section}}
\renewcommand{\theHequation}{supp.\arabic{equation}}
\renewcommand{\theHfigure}{supp.\arabic{figure}}
\renewcommand{\theHtable}{supp.\arabic{table}}

\section*{Supplementary Material}
\section{Method details}
\label{app:method_details}

\subsection{Residual attention implementation}

The local-attention graph has effective neighborhood size $k_i=\min(k,n_i-1)$ and is constructed once from $\mathbf{X}_i^0$. It is reused by all local-attention blocks. Edge supervision and evaluation use a fixed initial-space 32-neighbor graph independently of the local-attention neighborhood size. Local-attention and feed-forward outputs use residual connections with learnable LayerScale vectors~\cite{touvron2021cait} and DropPath~\cite{huang2016stochasticdepth}:
\begin{align}
\widetilde{\mathbf{H}}_i
&=
\mathbf{H}_i+
\operatorname{DropPath}\!\left(
\gamma_a\odot\operatorname{LocalSA}(\mathbf{H}_i)
\right),\\
\mathbf{H}_i
&\leftarrow
\widetilde{\mathbf{H}}_i+
\operatorname{DropPath}\!\left(
\gamma_f\odot\operatorname{FFN}(\widetilde{\mathbf{H}}_i)
\right).
\end{align}
The feed-forward block uses a gated SwiGLU-style structure~\cite{noam2020swiglu}.

Cross-object messages are incorporated through
\begin{equation}
\widetilde{\mathbf{H}}_i
=
\mathbf{H}_i+
\operatorname{DropPath}\!\left(
\gamma_c\odot W_o\mathbf{Z}_i^{\mathrm{cross}}
\right).
\end{equation}
Each cross-object update is followed by local attention and feed-forward refinement.

\section{Dataset construction}
\label{app:data_details}

\subsection{Simulation pipeline}

All datasets are generated in SI units with MuJoCo~\cite{todorov2012mujoco}. Rigid objects are represented by free-joint bodies, while the deformable object is represented by a MuJoCo flex body with vertex-level degrees of freedom. We use a $0.002\,\mathrm{s}$ time step, implicit integration, elliptic friction cones, viscosity $6\times10^{-4}$, friction-to-normal constraint impedance ratio $\texttt{impratio}=5$, 200 solver iterations, and tolerance $10^{-10}$. Contact parameters are $\mathrm{solref}=(0.005,1)$ and $\mathrm{solimp}=(0.90,0.97,0.02)$, with joint damping $0.01$. Flat, Bumpy, and Jenga use contact dimensionality $\texttt{condim}=6$, enabling sliding, torsional, and rolling friction, whereas Deformable uses $\texttt{condim}=3$, which includes tangential sliding friction but not torsional or rolling friction.

For Flat, Bumpy, and Deformable, gravity is sampled as $g_z=-9.81(1+\epsilon)$ with $\epsilon\sim\mathcal{U}[-0.05,0.05]$; Jenga uses fixed gravity $g_z=-9.81\,\mathrm{m/s^2}$. External forces are applied during the first $0.05\,\mathrm{s}$. Simulation terminates at $10\,\mathrm{s}$ for Flat, Bumpy, and Jenga and at $8\,\mathrm{s}$ for Deformable, or earlier when the maximum absolute generalized velocity remains below $10^{-3}$ for $0.2\,\mathrm{s}$.

Flat and Bumpy sample rigid-object materials uniformly from HDPE, rubber, pine, and oak. The Jenga impactors and the two Deformable impactors use the complete eight-material pool in Table~\ref{tab:material_parameters}, while the Jenga tower blocks are fixed to pine. Sliding, torsional, and rolling friction are independently perturbed by $\pm10\%$, $\pm20\%$, and $\pm20\%$, respectively, and clamped to $[0.12,0.60]$, $[0.005,0.020]$, and $[0.003,0.012]$. The supporting surface uses fixed HDPE parameters.

\begin{table*}[t]
\centering
\caption{MuJoCo rigid-geometry parameter presets. Density is given in $\mathrm{kg/m^3}$; sliding friction $\mu^t$ is dimensionless, while torsional and rolling coefficients $\mu^s$ and $\mu^r$ have units of length. The coefficients are simulation parameters intended to span diverse contact behavior rather than calibrated measurements of specific material pairs.}
\label{tab:material_parameters}
\begin{tabular}{lrr@{\quad}lrr}
\toprule
Material & $\rho$ $(\mathrm{kg/m^3})$ & $\bm{\mu}$ & Material & $\rho$ $(\mathrm{kg/m^3})$ & $\bm{\mu}$ \\
\midrule
HDPE & 950 & $(0.22,0.012,0.012)$ & Rubber & 1100 & $(0.55,0.020,0.012)$ \\
Pine & 500 & $(0.35,0.018,0.012)$ & Oak & 700 & $(0.40,0.020,0.012)$ \\
Ice & 917 & $(0.10,0.006,0.006)$ & Glass & 2500 & $(0.25,0.012,0.012)$ \\
Concrete & 2400 & $(0.50,0.020,0.012)$ & Aluminum & 2700 & $(0.35,0.018,0.012)$ \\
\bottomrule
\end{tabular}
\end{table*}

\subsection{Splits and preprocessing}

Training, validation, and test splits use the same scenario-specific sampling distributions but disjoint random-number streams. Validation and test generation additionally use split-specific random seeds for individual simulation attempts, preventing accepted samples from being shared across splits. Table~\ref{tab:dataset_statistics} summarizes the resulting split sizes and mesh resolutions.

\begin{table*}[t]
\centering
\caption{Dataset statistics for the four scenarios. Vertex counts are totals per sample, with object-wise composition in parentheses. Coarse meshes use a fixed subsampling stride of four.}
\label{tab:dataset_statistics}
\setlength{\tabcolsep}{3pt}
\begin{tabular}{lrrrrrr}
\toprule
Scenario & Objects & Train & Val. & Test & Fine vertices / sample & Coarse vertices / sample \\
\midrule
Flat & 9 & 30,000 & 1,000 & 1,000 & 5,778 $(9\times642)$ & 1,449 $(9\times161)$ \\
Bumpy & 9 & 30,000 & 1,000 & 1,000 & 5,778 $(9\times642)$ & 1,449 $(9\times161)$ \\
Jenga & 10 & 40,000 & 2,000 & 2,000 & 2,116 $(2\times642+8\times104)$ & 530 $(2\times161+8\times26)$ \\
Deformable & 3 & 5,000 & 200 & 200 & 1,577 $(2\times642+293)$ & 396 $(2\times161+74)$ \\
\bottomrule
\end{tabular}
\end{table*}

For the controlled comparison, the Flat split contains 1,000/100/100 training/validation/test scenes. Each trajectory is recorded every $0.008\,\mathrm{s}$ and terminates at $2\,\mathrm{s}$, yielding 250 physical transitions in addition to the two preceding frames required to initialize HOPNet. Scene identifiers, split membership, and underlying physical trajectories are shared. PIM and Small-PIM use separately preprocessed 642-vertex initial--terminal mesh pairs. HOPNet constructs its method-specific collision-complex representation and uses all 250,000 training and 25,000 validation transitions per epoch. FIGNet-Reimplemented and MeshGraphNet-Reimplemented independently construct their respective 64-node graphs and sample 16 transitions per training scene and 16 fixed transitions per validation scene, yielding 16,000 training and 1,600 validation examples per epoch.

\subsection{Scenario details}

Table~\ref{tab:scene_sampling} summarizes the initial-state distributions. Intervals denote independent uniform samples; the signs of the Flat/Bumpy corner velocities are selected to point toward the interaction region.

\begin{table*}[t]
\centering
\caption{Scenario-specific initial-state ranges shown component-wise. For Flat/Bumpy objects F--I, the signs of the planar velocities are selected to point toward the interaction region.}
\label{tab:scene_sampling}
\begin{tabular}{llccc}
\toprule
Objects & Axis
& Position $(\mathrm{m})$
& Velocity $(\mathrm{m/s})$
& Force $(\mathrm{N})$ \\
\midrule
Flat/Bumpy A
& $x$ & $[-.04,.04]$ & $[-.01,.01]$ & $[-6,6]$ \\
& $y$ & $[-.04,.04]$ & $[-.01,.01]$ & $[-6,6]$ \\
& $z$ & $[.10,.14]$ & $0$ & $0$ \\
\addlinespace

Flat/Bumpy B--E
& $x$ & $\pm[.125,.175]$ & $[-.01,.01]$ & $[-6,6]$ \\
& $y$ & $\pm[.125,.175]$ & $[-.01,.01]$ & $[-6,6]$ \\
& $z$ & $[.10,.14]$ & $0$ & $0$ \\
\addlinespace

Flat/Bumpy F--I
& $x$ & $\pm[.25,.4]$ & $\pm1.3$ & $[-6,6]$ \\
& $y$ & $\pm[.25,.4]$ & $\pm1.3$ & $[-6,6]$ \\
& $z$ & $[.10,.14]$ & $0$ & $0$ \\
\addlinespace

Jenga A
& $x$ & $[-.45,-.38]$ & $[2.8,3.5]$ & $[10,14]$ \\
& $y$ & $[-.075,-.045]$ & $[-.03,.03]$ & $[-.4,.4]$ \\
& $z$ & $[.05,.08]$ & $[.05,.25]$ & $[-.2,.2]$ \\
\addlinespace

Jenga B
& $x$ & $[.38,.45]$ & $[-3.5,-2.8]$ & $[-14,-10]$ \\
& $y$ & $[.045,.075]$ & $[-.03,.03]$ & $[-.4,.4]$ \\
& $z$ & $[.14,.17]$ & $[.2,.5]$ & $[-.2,.2]$ \\
\addlinespace

Deformable A
& $x$ & $[-.12,.12]$ & $[-.25,.25]$ & $[-1,1]$ \\
& $y$ & $[-.30,-.16]$ & $[.8,2.0]$ & $[2,8]$ \\
& $z$ & $[.05,.09]$ & $[-.1,.2]$ & $[-.5,.5]$ \\
\addlinespace

Deformable B
& $x$ & $[-.12,.12]$ & $[-.25,.25]$ & $[-1,1]$ \\
& $y$ & $[.16,.30]$ & $[-2.0,-.8]$ & $[-8,-2]$ \\
& $z$ & $[.05,.09]$ & $[-.1,.2]$ & $[-.5,.5]$ \\
\bottomrule
\end{tabular}
\end{table*}

\paragraph{Flat.}
Nine spheres of radius $0.05\,\mathrm{m}$ interact on a horizontal HDPE surface. Objects A--E initialize within or around the central interaction region, while F--I enter from four outer quadrants. Variations in material, planar force, gravity, and friction produce collision, sliding, rolling, and stabilization outcomes.

\paragraph{Bumpy.}
Bumpy retains the same sphere distributions but replaces the plane by a fixed $513\times513$ height field over $[-0.98,0.98]^2\,\mathrm{m}^2$. Its height ranges from $-0.01091$ to $0.01130\,\mathrm{m}$ and combines $0.01\,\mathrm{m}$ base perturbations with 12 pits of depth $0.005\,\mathrm{m}$ and a $0.04\,\mathrm{m}$ boundary fade. The terrain is generated once with a random seed. Flat and Bumpy use separate models.

\paragraph{Jenga.}
The tower contains eight stationary pine blocks in four alternating layers centered at heights $0.025$, $0.075$, $0.125$, and $0.175\,\mathrm{m}$. Each layer contains two blocks whose centers are offset by $\pm0.055\,\mathrm{m}$ along the horizontal axis perpendicular to the blocks' long direction. Alternating block dimensions are $0.280\times0.070\times0.050\,\mathrm{m}^3$ and $0.070\times0.240\times0.050\,\mathrm{m}^3$. Two radius-$0.05\,\mathrm{m}$ impactors, whose materials are sampled from Table~\ref{tab:material_parameters}, approach from opposite sides.

\paragraph{Deformable.}
Two radius-$0.05\,\mathrm{m}$ rigid spheres approach a stationary flex box from opposite sides. The box uses full vertex degrees of freedom on a $13\times6\times6$ lattice spanning $0.34\times0.14\times0.12\,\mathrm{m}^3$, with total mass $0.5\,\mathrm{kg}$, flex radius $0.009\,\mathrm{m}$, friction $(1.0,0.04,0.02)$, and edge constraints with $\mathrm{solref}=(0.022,1)$ and $\mathrm{solimp}=(0.92,0.98,0.01)$. Its exported surface contains 293 vertices and 580 triangular faces.

For Flat, Bumpy, and Jenga, an object pair is counted as impacted after a three-step warm-up when it is in contact with relative speed at least $0.03\,\mathrm{m/s}$ and either normal force at least $20\,\mathrm{N}$ or detected penetration. Flat retains scenes with at least eight of the 36 sphere pairs impacted, whereas Bumpy requires at least five impacted pairs. Jenga monitors the 16 impactor--block pairs and requires at least two impacted pairs, at least one impact involving each impactor, and tower collapse. Collapse requires at least three blocks to translate horizontally by more than $0.05\,\mathrm{m}$ or tilt by more than $30^\circ$, including a top-layer block, or a top-layer block center to fall below $0.10\,\mathrm{m}$. Non-finite Jenga states are rejected, as are samples in which a tower block exceeds $0.60\,\mathrm{m}$ total displacement, $0.55\,\mathrm{m}$ horizontal displacement, or the height interval $[-0.02,0.40]\,\mathrm{m}$.

Deformable candidates are retained only when at least one rigid sphere contacts the flex body and the maximum soft-vertex displacement reaches $0.002\,\mathrm{m}$; initial rigid--soft contact is rejected. These event-conditioned filters prevent the retained datasets from being dominated by non-interacting or numerically failed trials while preserving variation in terminal outcomes.

\paragraph{OOD input ranges.}
Table~\ref{tab:ood_ranges} summarizes the absolute impactor input ranges for Jenga and Deformable.

\begin{table}[!htb]
\centering
\caption{Absolute impactor input ranges. Lower/upper OOD correspond to close/far positions, slow/fast velocities, and low/high forces.}
\label{tab:ood_ranges}
\small
\setlength{\tabcolsep}{2.5pt}
\begin{tabular}{@{}lccc@{}}
\toprule
Factor & ID & Lower OOD & Upper OOD \\
\midrule
\multicolumn{4}{@{}l}{\textit{Jenga ($x$ axis)}} \\
Position (m) & $[0.38,0.45]$ & $[0.30,0.37]$ & $[0.46,0.53]$ \\
Velocity (m/s) & $[2.8,3.5]$ & $[2.0,2.7]$ & $[3.6,4.3]$ \\
Force (N) & $[10.0,14.0]$ & $[5.5,9.5]$ & $[14.5,18.5]$ \\
\addlinespace[2pt]
\multicolumn{4}{@{}l}{\textit{Deformable ($y$ axis)}} \\
Position (m) & $[0.16,0.30]$ & $[0.13,0.15]$ & $[0.31,0.38]$ \\
Velocity (m/s) & $[0.8,2.0]$ & $[0.4,0.7]$ & $[2.1,2.4]$ \\
Force (N) & $[2.0,8.0]$ & $[0.5,1.5]$ & $[8.5,10.0]$ \\
\bottomrule
\end{tabular}
\end{table}

\subsection{Jenga and Deformable generalization}
\label{app:jenga_generalization}

We evaluate the fixed validation-selected Jenga checkpoints without retraining or fine-tuning. Table~\ref{tab:ood_ranges} specifies the original and shifted absolute $x$-component ranges of the two oppositely directed impactors. Each single-factor initial-condition OOD set contains 2,000 accepted scenes, with 900 matched-protocol and 100 outcome-unfiltered scenes from each of the lower- and upper-range branches. The matched protocol retains the original impact and tower-collapse requirements. The outcome-unfiltered protocol removes only these two requirements while retaining the initial-overlap checks, numerical- and non-finite-state rejection, and the displacement and height bounds described above. ID-U contains 2,000 outcome-unfiltered scenes sampled from the original input ranges. For these sets, all non-varied coordinates, materials, friction parameters, gravity, tower geometry, simulation horizon, and the $0.05\,\mathrm{s}$ force duration remain unchanged.

The separate shape-OOD set contains 2,000 matched-protocol scenes and changes only top-layer object J from its training-time elongated block to an equal-volume cube with side length $0.094354\,\mathrm{m}$. Both the visual mesh and MuJoCo collision geometry are changed to the cube. The cube retains the original object's 104 vertices, 204 triangular faces, vertex order, and face-index order, preserving the ten-object model interface while changing its geometry. Object I, the impactors, all initial-condition ranges, materials, friction parameters, gravity, simulation horizon, force duration, and event-conditioned acceptance criteria remain unchanged.

Main-paper Figure~\ref{fig:jenga_ood_qualitative} shows matched-protocol close position, far position, and shape OOD comparisons.

\paragraph{Deformable initial-condition generalization.}
We evaluate fixed validation-selected Deformable checkpoints without retraining or fine-tuning on separate position, velocity, and force OOD sets. Each set contains 200 matched-protocol scenes, with 100 from each lower/upper branch. Only the selected factor's absolute $y$-component ranges change according to Table~\ref{tab:ood_ranges}; the two impactors retain opposite directions, and all other generation settings remain unchanged. Accepted scenes require rigid--soft contact and a maximum soft-vertex displacement of at least $0.002\,\mathrm{m}$ observed during simulation, while retaining initial-contact and numerical checks. No outcome-unfiltered or shape-OOD set is used for Deformable. We additionally report vertex, centroid, and KNN edge-length errors for the 293-vertex soft box separately from the scene-level metrics over all three objects (1,577 vertices).

\section{Training and baseline details}
\label{app:training_details}

\subsection{PIM training}

Flat and Bumpy use a hidden dimension of 512, eight attention heads, two intra-object layers, and two interaction layers; Jenga and Deformable use a hidden dimension of 768, six attention heads, four intra-object layers, and two interaction layers. All configurations use KNN size 32, 32 FPS source anchors, and a feed-forward expansion factor of four.
Coarse training on stride-four meshes is used as initialization rather than as inference-time multi-resolution feature fusion. After validation-based checkpoint selection, the best coarse checkpoint initializes the same architecture at fine resolution. The fine stage recomputes vertex features and KNN indices on the full-resolution meshes; coarse features are not concatenated with fine features or reused during fine-stage inference. Checkpoint selection in both stages uses validation coordinate L1 error.

All models use AdamW~\cite{loshchilov2019adamw} with linear warm-up followed by cosine decay. Flat and Bumpy use weight decay $3\times10^{-4}$ and dropout $0.1$, whereas Jenga and Deformable use weight decay $10^{-4}$ and no dropout. All configurations use DropPath rate $0.1$ and LayerScale initialization $10^{-3}$.

The Flat and Bumpy coarse stages train for 80 epochs with learning rate $3\times10^{-4}$, minimum learning rate $10^{-5}$, and a 4\% warm-up. The Jenga and Deformable coarse stages use the same learning-rate schedule for 100 epochs. Loss weights and the edge-weight ramp follow the main-paper objective~\eqref{eq:training_objective}.

Flat fine training runs for 40 epochs with learning rate $1.2\times10^{-4}$, minimum learning rate $10^{-5}$, and a 5\% warm-up. Bumpy fine training runs for 20 epochs with learning rate $2\times10^{-5}$, minimum learning rate $2\times10^{-6}$, and a 15\% warm-up. Jenga and Deformable fine training run for 40 epochs using the Flat fine-stage learning-rate and loss settings. Per-stage batch sizes and chunk sizes are listed in Table~\ref{tab:training_configuration}.

\begin{table*}[t]
\centering
\caption{Coarse and fine training configurations. Global batch size includes all samples processed across four GPUs and all gradient-accumulation steps.}
\label{tab:training_configuration}
\setlength{\tabcolsep}{4pt}
\begin{tabular}{llrrrrr}
\toprule
Scenario & Stage & Batch/GPU & Accum. & Global & Local chunk & Cross chunk \\
\midrule
Flat/Bumpy & Coarse & 16 & 1 & 64 & 256 & 16 \\
Flat/Bumpy & Fine & 4 & 2 & 32 & 1024 & 16 \\
Jenga/Deform. & Coarse & 32 & 1 & 128 & 256 & 16 \\
Jenga/Deform. & Fine & 8 & 1 & 32 & 1024 & 16 \\
\bottomrule
\end{tabular}
\end{table*}

Models are implemented in PyTorch and trained on four NVIDIA RTX A6000 GPUs using distributed data parallelism and mixed precision. 
All attention layers use PyTorch scaled dot-product attention with the FlashAttention backend~\cite{dao2022flashattention}. Gradient norm is clipped to 1.0. Local queries and object pairs are processed in chunks without changing neighborhoods or pairwise softmax normalization.

\subsection{Direct equilibrium-state baselines}

\paragraph{Rigid-Pose MLP.}
Rigid-Pose MLP represents each object by a 30-dimensional geometric and physical descriptor containing its initial center, center relative to the scene center, axis-aligned extent, the root-mean-square distance of vertices from the object centroid, six upper-triangular covariance entries, initial velocity, applied force, density, three friction coefficients, and four scene-level conditions. A learned 16-dimensional embedding of each object's index in the fixed object ordering is concatenated to this descriptor. A shared layer-normalized MLP maps each object to a 256-dimensional embedding. Mean and max pooling over the object embeddings are concatenated and projected to a 256-dimensional scene context, which is broadcast back to every object. The object and scene embeddings are then concatenated and processed by one residual MLP block of width 256.

Two prediction heads output a three-dimensional center displacement and a continuous 6D rotation representation~\cite{zhou2019continuity}. The latter is converted to a proper rotation matrix by orthogonalization. For object $i$, the predicted terminal vertices are
\begin{equation}
\widehat{\mathbf{Y}}_i
=
\left(\mathbf{X}_i^0-\mathbf{c}_i^0\right)\mathbf{R}_i
+
\mathbf{c}_i^0
+
\Delta\mathbf{c}_i.
\end{equation}
The final translation and rotation layers are initialized to the identity transformation. Rigid-Pose MLP therefore predicts a rigid terminal pose for every object. It is evaluated on all four scenarios; on Deformable, it serves as a constrained rigid-output reference and cannot reproduce local deformation.

\paragraph{Vertex Transformer.}
Vertex Transformer is a hierarchical vertex-level equilibrium-state predictor developed for this study, rather than a reimplementation of the Point Transformer architecture of Zhao et al.~\cite{zhao2021pointtransformer}. Each vertex receives centered coordinates, RMS-normalized coordinates, normalized radius, and three sinusoidal/cosine Fourier-coordinate bands, producing a 25-dimensional geometric feature. The same 30-dimensional object descriptor used by Rigid-Pose MLP is concatenated with a learned 16-dimensional embedding of the object's index in the fixed object ordering and projected to the model width. The projected condition is added to every vertex token of the corresponding object.

The network uses hidden width 384, six attention heads, KNN size 32, and feed-forward expansion factor four. One intra-object vertex-attention block applies six-head self-attention over cached KNN neighborhoods. Embedded relative coordinate differences are added to the neighbor key and value features and also parameterize a learned per-head attention bias. Mean and max pooling then reduce each object to one token, and one standard Transformer encoder layer exchanges information among the object tokens. The resulting object context is projected and broadcast back to all vertices of the corresponding object. A zero-initialized MLP directly predicts a three-dimensional displacement for each vertex:
\begin{equation}
\widehat{\mathbf{Y}}_i
=
\mathbf{X}_i^0
+
\operatorname{MLP}_{\mathrm{dec}}(\mathbf{H}_i).
\end{equation}
Unlike PIM, it exchanges inter-object information only through pooled object tokens and does not use point-level cross-object attention or the center--relative dual-branch decoder.

\paragraph{Training protocol.}

Both direct equilibrium-state prediction baselines operate on the fine-resolution meshes and use the same training, validation, and test splits and the same equilibrium-state evaluator as PIM. Each model is trained independently for 100 epochs on one NVIDIA RTX A6000 with mixed precision, AdamW, initial learning rate $10^{-4}$, weight decay $10^{-4}$, a 5\% linear warm-up, and cosine decay to $10^{-5}$. Gradients are clipped to norm 1.0. Training uses the main-paper objective~\eqref{eq:training_objective} with the direct-baseline weights; checkpoints are selected by validation coordinate L1 error.

\subsection{Controlled Flat comparison}

\paragraph{PIM variants.}
The controlled PIM retains the Flat architecture and objective used in the main experiments: hidden width 512, eight attention heads, two intra-object layers, two cross-object interaction layers, feed-forward expansion factor four, KNN size 32, and 32 source anchors. It has 19,755,622 trainable parameters. PIM is trained for 80 coarse epochs with initial learning rate $3\times10^{-4}$ and then for 40 fine epochs with initial learning rate $1.2\times10^{-4}$. Small-PIM retains the input representation, KNN and anchor counts, both interaction layers, decoder, losses, and the same 80/40-epoch coarse-to-fine schedule. It uses hidden width 256, four attention heads, three pre-interaction intra-object layers, two interaction layers, and feed-forward expansion factor three, giving 5,013,302 trainable parameters.

\paragraph{HOPNet.}
HOPNet~\cite{Wei2025} is retrained using its released architecture and objective, with 128 latent channels, one graph-processing layer, and two-layer ReLU MLPs. Rather than using PIM's preprocessed initial--terminal mesh pairs, it constructs its method-specific collision-complex representation from every recorded transition. The complexes use an object radius of $0.1\,\mathrm{m}$ and a $0.01\,\mathrm{m}$ local face neighborhood, and retain the scene-specific table properties. HOPNet is trained for 40 epochs with Adam, a constant learning rate of $10^{-4}$, and all 250,000 training transitions per epoch; the final-epoch checkpoint is used for evaluation. Its rollout consumes the required history and performs 249 learned updates to reach the exact $2\,\mathrm{s}$ target. HOPNet-Adapt appends the normalized external-force vector applied during the initial forcing interval to the object feature:
\begin{equation}
\mathbf{x}^{(4)}_{i,t,\mathrm{adapt}}
=
\left[\mathbf{x}^{(4)}_{i,t},\mathbf{f}_{i,t}/10\right]
\in\mathbb{R}^{16}.
\end{equation}
The force is zero for the static floor, and all other model, target, graph, and optimization settings remain unchanged.

\paragraph{FIGNet-Reimplemented.}
FIGNet-Reimplemented is based on the released PyTorch implementation of FIGNet~\cite{Yi_FIGNet_PyTorch_implementation_2024} while following the architecture and one-step objective described by the original method~\cite{allen2023fignet}. It is therefore our PyTorch reimplementation for the controlled Flat comparison rather than an execution of the official FIGNet source implementation. Each sphere is represented by 64 vertices, and the table is represented by explicit boxes. The model consumes three consecutive position frames, uses ten message-passing blocks of width 128 with two-layer ReLU MLPs, and predicts a normalized one-step position acceleration. Training uses Gaussian position noise with standard deviation $3\times10^{-5}\,\mathrm{m}$, gradient clipping at one, Adam with constant learning rate $10^{-4}$, and 100 epochs. Each epoch samples 16 transitions per training scene and uses 16 fixed transitions per validation scene. During rollout, sphere--sphere and sphere--box face interactions are reconstructed from the predicted state using a $0.01\,\mathrm{m}$ surface margin. A least-squares rigid transform is applied after every predicted update and is fed back into the next step. FIGNet-Reimplemented-Adapt changes only the physical-property vector from 9 to 12 dimensions by appending the raw three-dimensional external-force vector.

\paragraph{MeshGraphNet-Reimplemented.}
Our PyTorch Geometric implementation adapts the MeshGraphNets architecture~\cite{pfaff2021meshgnn} to three-dimensional rigid spheres. It is our reimplementation for the controlled Flat comparison rather than an execution of an official rigid-body implementation; the codebase from which it was developed originally targeted two-dimensional cylinder flow. It uses 64 vertices per sphere, 15 message-passing blocks of width 128, static triangle edges, and dynamic inter-object world edges with radius $0.03\,\mathrm{m}$ and at most eight senders per receiver. The network predicts a normalized one-step vertex-velocity increment, after which positions are updated by semi-implicit Euler with $\Delta t=0.008\,\mathrm{s}$. Training uses velocity noise with standard deviation $0.002\,\mathrm{m/s}$, Adam with constant learning rate $10^{-4}$, 100 epochs, and the same 16 training and 16 fixed validation transitions per scene used by FIGNet-Reimplemented. No rigidity constraint is applied during rollout. A proper rigid transform is fitted only to the terminal 64-vertex output to transfer it to the common 642-vertex evaluation mesh, and this terminal bridge is never fed back into the trajectory. MeshGraphNet-Reimplemented-Adapt changes only the node feature from 14 to 17 dimensions by appending $\mathbf{f}_{i,t}/10$.

The trainable parameter counts for HOPNet, HOPNet-Adapt, FIGNet-Reimplemented, FIGNet-Reimplemented-Adapt, MeshGraphNet-Reimplemented, and MeshGraphNet-Reimplemented-Adapt are 707,078, 707,462, 5,560,454, 5,561,222, 2,878,723, and 2,879,107, respectively.

\paragraph{Data and comparison protocol.}
All models are trained on one RTX A6000. The scene manifests, split membership, physical trajectories, force records, and terminal evaluation meshes are shared, but the training data are not converted into a single common representation: PIM uses initial--terminal mesh pairs, HOPNet uses collision complexes, and the two graph-network reimplementations construct their own temporal graphs and one-step targets. Consequently, the comparison controls the underlying simulated scenes and terminal targets, but not the learning objective, number of supervised transitions, parameter count, or optimization cost.

HOPNet-Adapt is the strongest rollout baseline with vertex error $0.038674\,\mathrm{m}$ and centroid error $0.037202\,\mathrm{m}$, followed by FIGNet-Reimplemented-Adapt at $0.043898\,\mathrm{m}$, HOPNet at $0.049890\,\mathrm{m}$, FIGNet-Reimplemented at $0.079846\,\mathrm{m}$, and the two MeshGraphNet variants in terms of vertex error. Adding the interval force reduces HOPNet vertex and centroid errors by $22.5\%$ and $23.3\%$, respectively. It reduces FIGNet vertex error by $45.0\%$, showing that explicit knowledge of the applied external force is useful for both architectures. The same modification increases MeshGraphNet vertex error by $10.8\%$, so providing the external-force input alone does not guarantee more accurate long-horizon rollout. The near-zero triangle-edge errors of HOPNet and FIGNet arise from their rigid pose output or per-step rigid projection, and those of MeshGraphNet arise from the terminal rigid bridge. They should not be interpreted as evidence that these models produce more accurate terminal configurations. PIM avoids autoregressive accumulation by predicting the terminal mesh directly. Despite using $74.6\%$ fewer parameters than PIM, Small-PIM increases vertex and centroid errors by only $4.1\%$ and $4.5\%$, respectively. At a parameter count comparable to FIGNet-Reimplemented-Adapt, it also achieves a $27.1\%$ lower vertex error, showing that the advantage of direct terminal prediction persists at substantially reduced model capacity.

\subsection{Controlled ablations}

The no-cross-attention ablation removes only the cross-object update while retaining local attention, feed-forward refinement, residual connections, and the dual-branch decoder. The direct-decoder ablation replaces the two decoder branches by a zero-initialized per-vertex MLP. Both use the full coarse-to-fine schedule, losses, splits, and checkpoint selection. The Deformable no-edge-loss ablation sets $\lambda_{\mathrm{edge}}=0$ in both the coarse and fine stages while retaining the architecture, data splits, coordinate and centroid losses, optimizer, training schedule, and validation-based checkpoint selection. Its fine stage is initialized from the corresponding no-edge-loss coarse checkpoint.

\subsection{Hyperparameter sensitivity}
\label{app:hyperparameter_sensitivity}

We conduct the hyperparameter sensitivity study on Deformable, whose coarse representation contains 161 vertices per rigid sphere and 74 vertices for the soft box. Consequently, all evaluated values in $\{16,32,64\}$ correspond to distinct neighborhoods and source-anchor sets. Starting from the default $(k,a)=(32,32)$, we vary one quantity at a time using $(16,32)$, $(64,32)$, $(32,16)$, and $(32,64)$. All variants retain the same architecture, parameter count, fixed data splits, coarse-to-fine schedule, optimizer, losses, and validation-based checkpoint criterion.

To decouple the local-attention neighborhood from geometric supervision, the edge loss and edge metric use a fixed 32-neighbor graph in every variant, while local attention uses the specified value of $k$. Each configuration is trained independently through the complete coarse-to-fine pipeline, and every fine-stage model is initialized from its matching coarse-stage checkpoint. The $(32,32)$ reference is retrained as part of the sweep rather than reusing the main-result checkpoint. We report test-set results from this matched run. Relative training time is computed from the median synchronized duration of 20 profiled fine-stage training steps in the first epoch on four RTX A6000 GPUs and normalized by the matched $(32,32)$ measurement.

\section{Evaluation details}
\label{app:evaluation_details}

\subsection{Geometric metrics}

The geometric errors and averaging conventions are defined in main-paper Eqs.~\eqref{eq:placement_errors} and~\eqref{eq:edge_errors}.

The controlled terminal comparison evaluates all eight models at the exact $2\,\mathrm{s}$ target on the same 100 test scenes and 642-vertex sphere meshes. PIM and Small-PIM make one direct equilibrium-state prediction, HOPNet performs 249 updates after consuming its required history, and the FIGNet and MeshGraphNet variants perform 250 updates from $t=0$. Model-specific history, state representation, graph construction, and rollout conventions are retained.

HOPNet predicts rigid poses, FIGNet applies a rigid projection after each update, and MeshGraphNet uses a terminal-only rigid resolution bridge. Their high-resolution triangle-edge errors therefore reflect the output representation rather than unconstrained shape accuracy and are excluded from ranking. Vertex and centroid metrics are computed per object and then per scene before equal scene averaging. All rollout methods produced finite outputs on all 100 test scenes. Under the evaluator's general failure policy, failed or non-finite scenes are retained in the failure count and excluded only from error statistics explicitly conditioned on successful scenes. Paired scene-level confidence intervals use 10,000 bootstrap resamples.

\subsection{Runtime protocol}
\label{app:runtime_details}

PIM inference is measured on an RTX A6000 with PyTorch 2.4.0 in eager mixed precision, compilation disabled, 20 warm-up passes, and 100 measured passes. Device-resident timing includes only the model forward pass. MuJoCo 3.3.7 runs on two Intel Xeon Gold 6330 CPUs. Its measured per-scene runtime includes state initialization, keyframe restoration, rollout, and final vertex extraction, excluding XML and asset compilation.

Flat, Bumpy, and Jenga use three warm-up scenes and 100 test scenes; Deformable uses one warm-up and ten scenes. Single-scene latency pairs model batch size one with one MuJoCo worker. Throughput pairs model batch size $B$ with $B$ concurrent MuJoCo workers and excludes process startup, XML compilation, and dispatch.

The controlled Flat timing uses the first two manifest-ordered test scenes for latency and the first 16 for throughput. PIM and Small-PIM use batch size one for latency and batch size eight for throughput, whereas each rollout model uses one independent worker for latency and eight independent workers for throughput, each with batch size one. The matched $2\,\mathrm{s}$ equilibrium-state protocol measures one direct equilibrium-state prediction for PIM, 249 updates for HOPNet, and 250 updates for FIGNet and MeshGraphNet to reach $2\,\mathrm{s}$. The supplementary 100/102-step rollout timing protocol excludes PIM and measures the released 102-update HOPNet rollout or a 100-update diagnostic for FIGNet and MeshGraphNet. Because the 102- and 100-update protocols are not identical, cross-family native-scale runtime ratios are descriptive rather than strictly task aligned.

Timing begins after checkpoint loading, scene loading, and static input preparation and ends after synchronized prediction. It includes required transfers, dynamic collision or interaction-graph construction, neural inference, autoregressive state updates, and rigid fitting where required, but excludes output serialization and error computation. Latency uses one worker and batch size one. Throughput uses batch size eight for PIM or eight concurrent rollout workers on the same RTX A6000. All models are measured sequentially with one OpenMP, MKL, OpenBLAS, and PyTorch CPU thread per process.

\subsection{Supplementary 100/102-step rollout timing}

\begin{table*}[t]
\centering
\caption{Supplementary rollout timing for the six autoregressive models. HOPNet uses its released 102-update setting, while the FIGNet and MeshGraphNet reimplementations use a 100-update diagnostic. Latency uses two scenes with one worker, and throughput uses 16 scenes and eight workers.}
\label{tab:native_runtime_results}
\setlength{\tabcolsep}{4pt}
\begin{tabular}{lrrr}
\toprule
Method & Calls & Latency (s) & Throughput (samples/s) \\
\midrule
HOPNet & 102 & 187.4733 & 0.021312 \\
HOPNet-Adapt & 102 & 168.3764 & 0.022649 \\
FIGNet-Reimpl. & 100 & 3.3253 & 0.645356 \\
FIGNet-Reimpl.-Adapt & 100 & 3.5275 & 0.637503 \\
MeshGraphNet-Reimpl. & 100 & \textbf{1.4982} & \textbf{1.985109} \\
MeshGraphNet-Reimpl.-Adapt & 100 & 1.6231 & 1.935324 \\
\bottomrule
\end{tabular}
\end{table*}

Table~\ref{tab:native_runtime_results} preserves the model-native rollout scale as a runtime-only diagnostic; no accuracy result is reported at this horizon. PIM is excluded because it has no corresponding released or native-scale trajectory protocol. Within each model family, the variants without the additional force input and their corresponding \emph{-Adapt} variants use identical call counts. HOPNet uses 102 updates, whereas FIGNet and MeshGraphNet use 100. Moreover, the original MeshGraphNets work does not define a rigid-body rollout protocol for this task. We therefore do not infer cross-family speedups from this table; the task-aligned $2\,\mathrm{s}$ terminal comparison remains the primary runtime result.

\section{Additional results}
\label{app:additional_results}

\subsection{Complete quantitative results}

Main-paper Table~\ref{tab:baseline_results} reports direct-predictor results and Jenga/Deformable generalization, including scene-level and object-specific vertex, centroid, and edge errors. Main-paper Table~\ref{tab:ablation_results} reports the architectural and edge-loss ablations, with qualitative edge-loss results in Figure~\ref{fig:no_edge_comparison}. Table~\ref{tab:hyperparameter_sensitivity} reports the local-neighborhood and source-anchor study.

Relative to the full model, removing cross-object attention increases vertex error by $135.1\%$, $98.3\%$, $120.5\%$, and $87.6\%$ on Flat, Bumpy, Jenga, and Deformable, respectively. The corresponding centroid-error increases are $180.6\%$, $109.2\%$, $202.4\%$, and $109.6\%$. The effect is therefore consistent across both rigid and deformable settings and is particularly pronounced for Jenga centroid prediction, where object placement depends on interactions throughout the stack. The lower edge errors obtained without cross-object attention on Flat, Bumpy, and Jenga do not contradict this conclusion: edge error measures local shape preservation and is insensitive to an incorrect global rigid pose.

The direct decoder produces higher vertex and centroid errors than the full dual-branch decoder in every scenario. Its vertex-error increases are $20.8\%$, $3.2\%$, $8.1\%$, and $1.3\%$ on Flat, Bumpy, Jenga, and Deformable, respectively. Thus, explicitly separating object-level translation from zero-mean relative vertex motion reduces vertex and centroid errors in all four evaluated single-run comparisons for terminal placement, but the small differences on Bumpy and Deformable indicate that the benefit is not uniformly large. On Jenga, the direct decoder reduces edge error by $29.8\%$ while increasing vertex and centroid errors by $8.1\%$ and $4.3\%$, again demonstrating that local geometric preservation and terminal placement evaluate different properties.

Removing the edge loss on Deformable increases vertex and centroid errors by $4.2\%$ and $5.7\%$, respectively, but increases the KNN edge-length error by $227.2\%$. The edge error therefore detects a loss of local geometric fidelity that is weakly reflected by aggregate pointwise placement metrics. Figure~\ref{fig:no_edge_comparison} shows the corresponding visible distortions, particularly on the nominally rigid spheres, illustrating that KNN edge-length supervision helps preserve local object geometry while still allowing the soft box to deform toward the ground-truth terminal geometry.

\begin{table*}[t]
\centering
\caption{Deformable hyperparameter sensitivity in matched single-run experiments using seed 1. Errors are test-set averages in meters and lower is better. Relative time is normalized by the matched $(k,a)=(32,32)$ reference. The edge neighborhood is fixed at 32, and all variants have the same number of learnable parameters.}
\label{tab:hyperparameter_sensitivity}
\setlength{\tabcolsep}{3.5pt}
\begin{tabular}{lcrrrr}
\toprule
Variant & $(k,a)$
& $\varepsilon_{\mathrm{vtx}}$
& $\varepsilon_{\mathrm{cen}}$
& $\varepsilon_{\mathrm{edge}}$
& Relative step time \\
\midrule
Smaller KNN & $(16,32)$ & 0.019769 & 0.013637 & $3.825\times10^{-4}$ & 0.783 \\
Default (sweep) & $(32,32)$ & 0.019539 & 0.013430 & $3.807\times10^{-4}$ & 1.000 \\
Larger KNN & $(64,32)$ & 0.019494 & 0.013441 & $3.866\times10^{-4}$ & 1.454 \\
Fewer anchors & $(32,16)$ & 0.019583 & 0.013454 & $3.780\times10^{-4}$ & 0.987 \\
More anchors & $(32,64)$ & 0.019636 & 0.013503 & $3.799\times10^{-4}$ & 0.988 \\
\bottomrule
\end{tabular}
\end{table*}

At fixed $a=32$, reducing the neighborhood to $k=16$ lowers fine-stage step time by $21.7\%$ while increasing vertex error by $1.18\%$. Increasing the neighborhood to $k=64$ raises step time by $45.4\%$, reduces vertex error by only $0.23\%$, and increases edge error by $1.55\%$. At fixed $k=32$, reducing the anchor count to $a=16$ increases vertex and centroid errors by $0.23\%$ and $0.18\%$, respectively, while reducing edge error by $0.71\%$. Increasing the count to $a=64$ increases vertex and centroid errors by $0.50\%$ and $0.54\%$, respectively, while reducing edge error by $0.21\%$. The measured step times for both alternatives remain within $1.3\%$ of the default. Thus, $a=32$ provides the best terminal-placement accuracy in this sweep, whereas fewer anchors provide only a small edge-preservation improvement.

\end{document}